\documentclass[aps,prb,twocolumn,superscriptaddress,longbibliography,nofootinbib]{revtex4-2}

\usepackage[utf8]{inputenc}
\usepackage[T1]{fontenc}

\usepackage{amssymb,amsmath,amsthm,bbm,bbold}
\usepackage{amsfonts}
\usepackage{graphicx}
\usepackage{mathtools}
\usepackage{hyperref}
\usepackage[normalem]{ulem}
\usepackage{color}
\usepackage{xcolor}
\usepackage{tcolorbox}

\usepackage{pifont}
\usepackage{amssymb}
\usepackage{tikz}
\usepackage{makecell}

\usepackage{pgf}

\newcommand{\Tr}{\mathrm{Tr}}

\newcommand{\HN}{\mathcal{H}_N}
\newcommand{\HNP}{\mathcal{H}_N^{(P)}}

\newcommand{\Fp}{\mathcal{F}} 

\newcommand{\FpP}{{\mathcal{F}^{(P)}}}

\newcommand{\dom}{\mathrm{dom}}

\newcommand{\ForcePrefactor}{\mathcal{G}}

\newcommand{\nalpha}{\vec n^{(\alpha)}}
\newcommand{\nbeta}{\vec n^{(\beta)}}

\newcommand{\bra}[1]{\mbox{$\langle #1 |$}}
\newcommand{\ket}[1]{\mbox{$| #1 \rangle$}}

\newcommand{\VFactor}[1]{L_{#1}}

\newtheorem{thm}{Theorem}

\newtheorem*{propo*}{Proposition}

\usepackage{float}
\usepackage{diagbox}
\usepackage{makecell}
\usepackage{lipsum}

\usepackage{ytableau}
\usepackage{braket}
\usepackage{tikz-cd}

\usepackage{enumitem}

\usepackage[caption=false]{subfig}

\newif\ifWeWantToUsePastTenseInConclusionSection

\newcounter{whichBECForceDiagramToUse}

\WeWantToUsePastTenseInConclusionSectiontrue

\begin{document}
\title{A Geometric Approach to Strongly Correlated Bosons: From $N$-Representability to the Generalized BEC Force}

\author{Chih-Chun Wang}
\affiliation{%
	Department of Physics, Arnold Sommerfeld Center for Theoretical Physics,
	Ludwig-Maximilians-Universit\"at M\"unchen,
	Theresienstr.~37, 80333 Munich, Germany
}
\affiliation{%
	Munich Center for Quantum Science and Technology (MCQST),
	Schellingstr.~4, 80799 Munich, Germany
}

\author{Christian Schilling}
\email{c.schilling@lmu.de}
\affiliation{%
	Department of Physics, Arnold Sommerfeld Center for Theoretical Physics,
	Ludwig-Maximilians-Universit\"at M\"unchen,
	Theresienstr.~37, 80333 Munich, Germany
}
\affiliation{%
	Munich Center for Quantum Science and Technology (MCQST),
	Schellingstr.~4, 80799 Munich, Germany
}

\date{\today}

\begin{abstract}
Building on recent advances in reduced density matrix theory, we develop a
  geometric framework for describing strongly correlated lattice bosons. We
  first establish that translational symmetry, together with a fixed pair
  interaction, enables an exact functional formulation expressed solely in
  terms of momentum occupation numbers. Employing the constrained-search
  formalism and exploiting a geometric correspondence between $N$-boson
  configuration states and their one-particle reduced density matrices, we
  derive the general form of the ground-state functional. Its structure
  highlights the omnipresent significance of one-body $N$-representability: (i)
  the domain is exactly determined by the $N$-representability conditions; (ii)
  at its boundary, the gradient of the functional diverges repulsively, thereby
  generalizing the recently discovered Bose-Einstein condensate (BEC) force;
  and (iii) an explicit expression for this boundary force follows directly
  from geometric arguments. These key results are demonstrated analytically for
  few-site lattice systems, and we illustrate the broader significance of our
  functional form in defining a systematic hierarchy of functional
  approximations.
\end{abstract}

\maketitle

\section{Introduction}\label{sec:intro}

Describing quantum many-body systems remains one of the central challenges in
physics, chemistry, and materials science. The exponential growth of the
Hilbert space with particle number renders exact wave-function-based
approaches intractable beyond very small systems, while the wave function
itself contains far more information than is typically required. Density
functional theory
(DFT)~\cite{hohenbergInhomogeneousElectronGas1964,kohnSelfconsistentEquationsIncluding1965}
circumvents this difficulty by replacing the wave function with the particle
density $n(\vec r)$ as the primary variable and has become an indispensable
\textit{ab initio} method in electronic structure
theory~\cite{burkePerspectiveDensityFunctional2012,beckePerspectiveFiftyYears2014,hasnipDensityFunctionalTheory2014}.
However, DFT provides only indirect access to correlation effects and therefore
often lacks predictive power for strongly correlated systems.

One-particle reduced density matrix functional theory (RDMFT) offers a
conceptually richer framework by using the full one-particle reduced density
matrix (1RDM) $\hat{\gamma}$ as its fundamental
variable~\cite{gilbertHohenbergKohnTheoremNonlocal1975,donnellyElementaryPropertiesEnergy1978,pernalReducedDensityMatrix2015}.
The 1RDM yields the kinetic energy exactly and reflects correlation directly through 
fractional occupation numbers, making RDMFT particularly suitable for the
description of strongly correlated systems. Over the past decades, RDMFT has
undergone substantial development through advances in its theoretical
foundations and scope~\cite{gilbertHohenbergKohnTheoremNonlocal1975,donnellyElementaryPropertiesEnergy1978,zumbachDensitymatrixFunctionalTheory1985,
pernalReducedDensityMatrix2015,schadeReducedDensitymatrixFunctionals2017,schillingCommunicationRelatingPure2018,giesbertzOnebodyReducedDensitymatrix2019,
schillingEnsembleReducedDensity2021,gibneyDensityFunctionalTheory2022,liebertReducedDensityMatrix2022,
liebertFoundationOneparticleReduced2022,rodriguez2022relativistic,liebert2023exact,liebertRefiningRelatingFundamentals2023,liebertReducedDensityMatrix2025,fredheimReducedDensityMatrix2025},
the understanding of general properties of the universal functional
\cite{schillingDivergingExchangeForce2019,cioslowskiBilinearConstraintsCorrelation2019,maciazekRepulsivelyDivergingGradient2021}, the derivation of exact functionals for
small systems
\cite{lopez-sandovalDensitymatrixFunctionalTheory2000,towsDensitymatrixFunctionalTheory2013,cohenLandscapeExactEnergy2016,benavides-riverosReducedDensityMatrix2020,liebertRefiningRelatingFundamentals2023},
the construction of approximate
functionals~\cite{mullerExplicitApproximateRelation1984,goedeckerNaturalOrbitalFunctional1998,yasudaCorrelationEnergyFunctional2001,
buijseApproximateExchangecorrelationHole2002,lopez-sandovalDensitymatrixFunctionalTheory2002,
kollmarNewApproachDensity2003,
lopez-sandovalInteractionenergyFunctionalLattice2004,gritsenkoImprovedDensityMatrix2005,pirisNaturalOrbitalFunctional2007,pirisNaturalOrbitalFunctional2011,
towsSpinpolarizedDensitymatrixFunctional2012,mitxelenaPerformanceNaturalOrbital2017,
benavides-riverosStaticCorrelatedFunctionals2018,giesbertzApproximateEnergyFunctionals2018,
vanmeerNonJKLDensityMatrix2018,
pirisNaturalOrbitalFunctional2019,benavides-riverosReducedDensityMatrix2020,
senjeanReducedDensityMatrix2022,
disabatinoIntroducingScreeningOnebody2022,
wangSelfconsistentfieldMethodCorrelated2022,irimiaSelfconsistentfieldMethodCorrelation2023,liebertFAes2023,
pirisAdvancesApproximateNatural2024},
and increasingly sophisticated algorithmic implementations~\cite{sharmaReducedDensityMatrix2008,lathiotakisDiscontinuitiesChemicalPotential2010,lemkeEfficientIntegraldirectMethods2022,cartierExploitingHessianBetter2024,yaoEnhancingReducedDensity2024,vladajVariationalMinimizationScheme2024}.

Despite these advances, the full potential of RDMFT remains unrealized. In particular, a comprehensive understanding of the structure of the universal functional, the consistent incorporation of $N$-representability, and the systematic construction of controlled approximations is still lacking.
This gap is especially pronounced for bosonic systems: although interacting bosons play a central role in
modern quantum many-body physics \cite{blochManybodyPhysicsUltracold2008}, bosonic RDMFT has only very recently been
established at a formal level
\cite{benavides-riverosReducedDensityMatrix2020,liebertReducedDensityMatrix2022}. In this context, RDMFT would be
particularly well suited, as it provides direct access to the momentum
density, determined through the time-of-flight method in experiments
\cite{andersonObservationBoseEinsteinCondensation1995}, and
allows for a universal characterization of Bose-Einstein condensation through
the Penrose-Onsager criterion, applicable to both homogeneous and
inhomogeneous systems~\cite{penroseBoseEinsteinCondensationLiquid1956}. It is precisely in this broader context that ultracold atomic gases have emerged as one of the central platforms of modern quantum many-body physics, providing an exceptionally clean and controllable platform
for exploring these
phenomena~\cite{andersonObservationBoseEinsteinCondensation1995,bradleyEvidenceBoseEinsteinCondensation1995,davisBoseEinsteinCondensationGas1995,blochUltracoldQuantumGases2005,blochManybodyPhysicsUltracold2008,blochQuantumSimulationsUltracold2012,grossQuantumSimulationsUltracold2017,schaferToolsQuantumSimulation2020},
with optical lattice realizations enabling controlled studies of
superfluid-Mott insulator
physics~\cite{jakschColdBosonicAtoms1998,jakschColdAtomHubbard2005a,greinerQuantumPhaseTransition2002,blochSuperfluidtoMottInsulatorTransition2022}.

\begin{figure}[htb]
  \includegraphics[width=1.02\columnwidth]{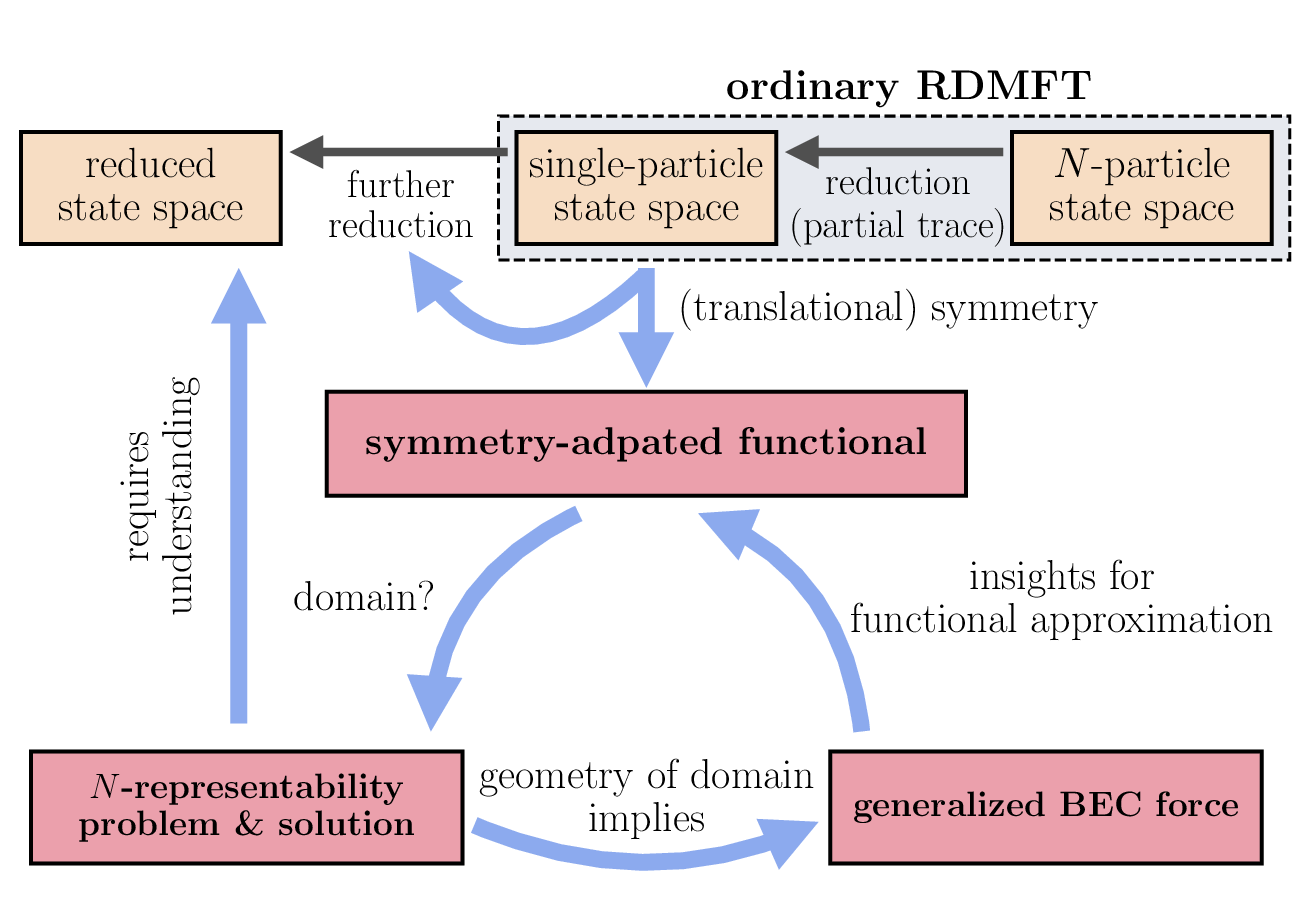}
  \caption{Conceptual overview of the symmetry-adapted functional-theoretical framework developed in this work. See text for details.}
  \label{fig:overview}
\end{figure}

In this work, we close this gap by showing that spatial symmetry,
$N$-representability, and the geometry of quantum states jointly determine the
structure of the universal interaction functional for lattice bosons. As
anticipated in Fig.~\ref{fig:overview}, incorporating translational symmetry
into RDMFT leads to a decisive simplification: the functional variable reduces
to the momentum occupation number vector. We demonstrate that the geometry of
the functional domain governs the functional form and enforces a universal
boundary behavior, namely a repulsive divergence of the functional gradient. This generalized BEC force extends earlier results from the BEC vertex to arbitrary boundary points, is derived here in closed analytical form, and reveals how
$N$-representability and state-space geometry constrain all admissible
functionals, providing a principled route toward their systematic
approximation.

This paper is organized as follows. In Sec.~\ref{sec:notation}, we review the general framework of RDMFT and explain how translational symmetry simplifies the functional domain. Section~\ref{sec:domain} presents a complete characterization of this domain, with an emphasis on small lattice systems. In Sec.~\ref{sec:F}, we derive the general form of the ground-state functional and analyze its geometric structure. Section~\ref{sec:BECforce} reveals the generalized BEC force, while Sec.~\ref{sec:ex} illustrates our findings for a Bose-Hubbard model and outlines a geometrically guided strategy for constructing functional approximations.

\section{Notation and Concepts}\label{sec:notation}

\subsection{Recap and Rationalization of RDMFT}
\label{sec:notation_general_summary}

Although bosonic systems are our primary focus, we briefly review in this section the basic elements of one-particle reduced density-matrix functional theory (RDMFT) for both bosonic and fermionic systems. This broader perspective enables a transparent comparison between the two cases. In particular, it allows us to identify structural features of functional theories that are distinctive to bosons and either absent or much less pronounced in the fermionic setting, where functional approaches are comparatively more mature and well established.

Throughout this paper, $\mathcal{H}_1$ and $\mathcal{H}_N$ denote the single-particle and $N$-particle Hilbert spaces, respectively. The many-particle Hilbert space is given by the symmetrized tensor product $\mathcal{H}_N=\mathrm{Sym}^N\mathcal{H}_1$ for bosons and by the antisymmetrized tensor product $\mathcal{H}_N=\bigwedge^N\mathcal{H}_1$ for fermions, reflecting the corresponding symmetry under particle exchange.

Functional theories are particularly effective because they do not target the
ground state of a single, isolated system, but rather provide a unified
description of an entire class of systems that share a common physical
structure \cite{liebertRefiningRelatingFundamentals2023}. In many-body physics and quantum chemistry, such classes naturally
arise through experimentally or physically controllable parameters. In
electronic-structure theory, for instance, the Hamiltonian depends on the
external potential $v(\vec r)$ determined by the nuclear configuration, while
in ultracold-atom experiments the parameters of the trapping potential, and
hence the one-particle term of lattice models such as the Bose-Hubbard
Hamiltonian, can be tuned
\cite{jakschColdBosonicAtoms1998,jakschColdAtomHubbard2005a}. Functional
theories take advantage of this shared structure to treat all members of the
class on equal footing.

This observation motivates the introduction of the concept of \emph{scope},
which constitutes one of the key conceptual ingredients of the present work.
The scope specifies the class of Hamiltonians to which a given functional
theory applies and thereby clarifies, in a systematic and conceptually
transparent manner, what the theory is designed to describe and what it can be
expected to deliver.

\begin{figure}
  \includegraphics[width=.78\columnwidth]{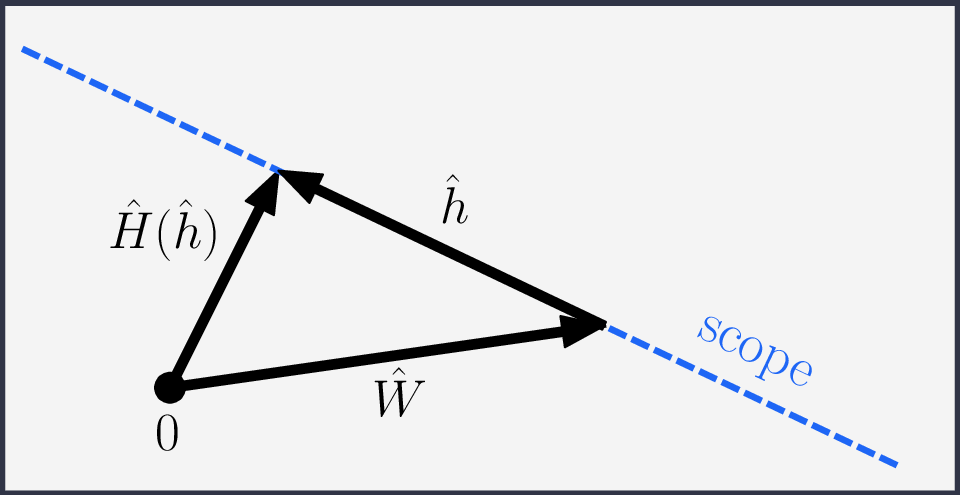}
  \caption{Schematic illustration of the concept of \emph{scope}: a functional theory is associated with a family of Hamiltonians $\hat H(\hat h)=\hat h+\hat W$, where $\hat h$ varies within a physically relevant real vector space of Hermitian operators determined by the problem class and its symmetries. The set of admissible Hamiltonians $\{\hat H(\hat h)\}$ thus forms an affine space (blue dashed line), referred to as the scope of the functional theory.}
  \label{fig:scope}
\end{figure}

In RDMFT, one considers Hamiltonians of the form
\begin{equation}
  \hat H(\hat h)=\hat h+\hat W,
\end{equation}
where $\hat h$ denotes a general one-particle operator, typically comprising
kinetic and external potential terms, and $\hat W$ is a fixed interaction
operator. Depending on the physical context, $\hat W$ may represent a contact
interaction in ultracold atomic gases, a Hubbard-type on-site interaction in
solid-state models, or the Coulomb interaction in quantum chemistry. The
admissible Hamiltonians $\{\hat H(\hat h)\}$ therefore form an affine subspace
of the space of Hermitian operators on $\mathcal{H}_N$, as illustrated
schematically in Fig.~\ref{fig:scope}. Specifying this affine space, i.e.\ the
scope, is the first essential ingredient of the functional theory.

The second ingredient is a variational principle. For ground-state problems,
this role is played by the Rayleigh-Ritz variational principle, which yields
the ground-state energy
\begin{equation}
  E(\hat h)=\min_{\ket{\Psi}}\braket{\Psi|\hat h+\hat W|\Psi}.
\end{equation}
Exploiting the common splitting $\hat H(\hat h)=\hat h+\hat W$, the expectation value separates naturally into a one-particle and an interaction contribution. Introducing the one-particle reduced density operator (1RDM) associated with an $N$-particle mixed or pure state $\hat\Gamma$,
\begin{equation}
  \label{eq:1rdm_from_N_particle_state}
  \hat\gamma \equiv N\,\Tr_{N-1}(\hat\Gamma),
\end{equation}
one may rewrite the variational problem in terms of $\hat\gamma$ and arrive at the constrained-search formulation of RDMFT \cite{levyUniversalVariationalFunctionals1979,valoneConsequencesExtending1matrix1980,liebertRefiningRelatingFundamentals2023}. In particular, the \emph{pure-state constrained-search functional} \cite{levyUniversalVariationalFunctionals1979}
\begin{equation}
  \label{eq:levylieb}
  \Fp[\hat\gamma]\equiv \min_{\ket{\Psi}\mapsto\hat\gamma}\braket{\Psi|\hat W|\Psi}
\end{equation}
encodes the interaction contribution universally across the entire scope. The
ground-state energy corresponding to any Hamiltonian $\hat H(\hat h)$ within
the scope is then obtained by minimizing $\Tr(\hat
h\hat\gamma)+\Fp[\hat\gamma]$ over all admissible $\hat\gamma$, a result that
follows directly from the Rayleigh-Ritz variational principle
\cite{gilbertHohenbergKohnTheoremNonlocal1975,levyUniversalVariationalFunctionals1979,valoneConsequencesExtending1matrix1980,liebertRefiningRelatingFundamentals2023}.

This construction makes explicit a central message of the present work:
functional theories arise from the interplay of two fundamental ingredients: the
specification of a scope and the choice of a variational principle. Once these
ingredients are fixed, the structure of the functional theory follows in a
systematic and largely theory-independent manner.

The minimization in Eq.~\eqref{eq:levylieb} is well defined only for one-particle density operators $\hat\gamma$ that originate from pure $N$-particle states, a condition known as \emph{(pure-state) $N$-representability}. The domain of the pure functional $\Fp$ is thus the set of all pure-state $N$-representable 1RDMs. For fermions, this set is restricted by the generalized Pauli constraints on the natural occupation numbers
\cite{borlandConditionsOnematrixThreebody1972,klyachkoQuantumMarginalProblem2006,
altunbulakPauliPrincipleRevisited2008,altunbulakPauliPrincipleRepresentation2008}.
For bosons, by contrast, every one-particle density operator with trace $N$ is pure-state $N$-representable
\cite{maciazekImplicationsPinnedOccupation2020,benavides-riverosReducedDensityMatrix2020},
which substantially simplifies the formulation and use of bosonic RDMFT.

From the scope-based perspective outlined above, it becomes transparent how
functional theories can be extended beyond ground states to target excited
states. Since the scope, i.e., the class of admissible Hamiltonians, remains unchanged,
such extensions are achieved by simply modifying the second fundamental
ingredient of the theory, namely the variational principle. Replacing the
Rayleigh-Ritz principle by an ensemble variational principle based on weighted
energies $E_{\boldsymbol{w}}\equiv\sum_i w_i E_i$ with non-increasing auxiliary
weights $\boldsymbol{w}$ \cite{grossDensityfunctionalTheoryEnsembles1988,dingGroundExcitedStates2024} yields ensemble
functionals that describe low-lying excitations. This approach underlies the
formulation of $\boldsymbol{w}$-ensemble RDMFT
\cite{schillingEnsembleReducedDensity2021,liebertFoundationOneparticleReduced2022}  and closely parallels developments
in ensemble density functional theory
\cite{theophilouEnergyDensityFunctional1979,GOK88b,GOK88c, Loos2020-EDFA, Fromager2020-DD, GK21, Cernatic21, GL23, SKCPJB24, GP24, Fromager24-EDFT, CLSF24}.
As a consequence, many of the conceptual insights and practical results derived
in this work can be obtained in a closely analogous manner within ensemble
formulations beyond the ground-state setting.

\subsection{Incorporating Translational Symmetry}
\label{sec:notation_translation_symmetry}

In the context of the present work—motivated by translationally invariant
lattice models for bosonic quantum matter—it is advantageous to incorporate
translational symmetry explicitly into the functional framework. More
generally, a systematic scope-based approach to incorporating symmetries into
functional theories has been developed at an abstract level and explicitly
realized for spin symmetries in Refs.~\cite{LMS26a,LMS26b}. Here, we demonstrate that
this framework can be adapted to the equally important case of translational
symmetry. As we will show, translational invariance has two key consequences:
it induces a decomposition of the many-particle Hilbert space into independent
symmetry sectors that can be addressed individually within a
functional-theoretical framework, and it allows for a simplification of both
the functional variable and its domain. Exploiting these aspects, we construct
a symmetry-adapted version of RDMFT tailored to translationally invariant
systems.

We consider $N$ bosons on a one-dimensional lattice with $d$ sites and periodic boundary conditions. Translational invariance implies that the Hamiltonian commutes with lattice translations, making the momentum representation particularly convenient. Let $\{\ket{k}\}_{k=0}^{d-1}$ denote the single-particle momentum eigenstates. An orthonormal basis of the $N$-particle Hilbert space is then given by the bosonic occupation-number states
\begin{equation}
  \label{eq:bosonic_fock_state}
  \ket{n_0,\dots,n_{d-1}}
  = \prod_{k=0}^{d-1}\frac{(\hat b_k^\dagger)^{n_k}}{\sqrt{n_k!}}\ket{0},
\end{equation}
with fixed particle number $\sum_{k=0}^{d-1} n_k = N$, where $\hat b_k^\dagger$ creates a boson with momentum $k$ and $\ket{0}$ denotes the vacuum state.

Each such configuration is an eigenstate of the translation operator $\hat g$, with eigenvalue determined by the total momentum,
\begin{equation}
  \label{eq:translation_g_eigenvalue}
  \hat g\ket{n_0,\dots,n_{d-1}}
  = e^{\frac{2\pi i}{d}\sum_{k=0}^{d-1} k n_k}\ket{n_0,\dots,n_{d-1}}.
\end{equation}
As a result, the $N$-particle Hilbert space decomposes into $d$ symmetry sectors labeled by the conserved total momentum $P=0,1,\dots,d-1$,
\begin{equation}\label{HNdecom}
  \mathcal{H}_N = \bigoplus_{P=0}^{d-1} \mathcal{H}_N^{(P)},
\end{equation}
with
\begin{equation}\label{HNP}
  \mathcal{H}_N^{(P)} =
  \mathrm{span}_{\mathbb{C}}
  \left\{\ket{n_0,\dots,n_{d-1}} \,\Big|\, \sum_{k=0}^{d-1} k n_k \equiv P \!\!\!\!\pmod d \right\}.
\end{equation}
As a simple illustration, for $N=3$ bosons on $d=3$ lattice sites the sectors $P=0,1,2$ are spanned by
$\{\ket{3,0,0},\ket{0,3,0},\ket{0,0,3},\ket{1,1,1}\}$,
$\{\ket{2,1,0},\ket{0,2,1},\ket{1,0,2}\}$, and
$\{\ket{1,2,0},\ket{0,1,2},\ket{2,0,1}\}$, respectively.

Since the Hamiltonian is block-diagonal with respect to the decomposition~\eqref{HNdecom}, variational ground-state methods can be applied separately in each symmetry sector $\mathcal{H}_N^{(P)}$. This sector-wise formulation yields not only the global ground state as the minimum over all $P$, but also the energetic minimum within each individual sector, while significantly reducing numerical cost. More importantly, in functional-theoretical frameworks it is essential: mixing different symmetry sectors leads to symmetry contamination, a well-known and severe pathology of practical functional approximations, most prominently encountered in the treatment of spin symmetries.

The second step in constructing the symmetry-adapted functional theory exploits the restriction imposed by translational invariance on the one-particle Hamiltonian. In this setting, the one-particle operator $\hat h$ reduces to a generalized kinetic-energy operator $\hat t$ that is diagonal in the momentum representation,
\begin{equation}
  \hat h \equiv \hat t = \sum_{k=0}^{d-1} t_k\, \hat b_k^\dagger \hat b_k .
\end{equation}
Here, the coefficients $t_k$ characterize a general single-particle dispersion
relation whose rate and range can be tuned in ultracold atom experiments \cite{Guenter2013,Schempp2015}, allowing for a broad class of effective kinetic energy operators. 
As a consequence, the one-particle energy
\begin{equation}
  \Tr(\hat h\hat\gamma)=\vec t\cdot\vec n
\end{equation}
depends only on the momentum occupation numbers
\begin{equation}\label{nk}
n_k \equiv \langle \Psi | \hat b_k^\dagger \hat b_k | \Psi \rangle = \braket{k|\hat\gamma|k},
\end{equation}
which form the vector $\vec n=(n_0,\dots,n_{d-1})^\top$, while $\vec
t=(t_0,\dots,t_{d-1})^\top$ collects the corresponding single-particle energies.
Thus, in the presence of translational symmetry, the full 1RDM contains
redundant information for the variational problem.

These observations lead directly to the introduction of a symmetry-adapted pure functional in a fixed momentum sector $P$,
\begin{equation}
  \label{eq:symmetry_sector_functional_def}
  \Fp^{(P)}[\vec n] \equiv
  \min_{\mathcal{H}_N^{(P)}\ni\ket{\Psi}\mapsto \vec n}
  \braket{\Psi|\hat W|\Psi},
\end{equation}
where $\ket{\Psi}\mapsto\vec n$ denotes the constraint
$\braket{\Psi|\hat b_k^\dagger \hat b_k|\Psi}=n_k$ for all $k$.
The domain of $\Fp^{(P)}$ consists of all momentum-occupation vectors $\vec n$ that are compatible with a pure $N$-boson state in the sector $\mathcal{H}_N^{(P)}$; this representability problem is addressed in Sec.~\ref{sec:domain}.

With the definition~\eqref{eq:symmetry_sector_functional_def}, the ground-state energy of $\hat h+\hat W$ within a fixed symmetry sector $\mathcal{H}_N^{(P)}$ is obtained as
\begin{equation}
  E^{(P)} = \min_{\vec n\in\dom\Fp^{(P)}}\bigl(\vec t\cdot\vec n+\Fp^{(P)}[\vec n]\bigr),
\end{equation}
in direct analogy to standard RDMFT. For each $P$, the functional $\Fp^{(P)}$ is universal in the sense that it depends only on the interaction $\hat W$ and not on the specific kinetic-energy operator. This universality highlights the economical nature of functional theories: once approximations to $\Fp^{(P)}$ are developed for a given interaction, they can be reused across an entire class of systems, in stark contrast to wave function-based approaches.

Finally, we note that translational invariance of the interaction operator $\hat W$ implies that it is likewise block-diagonal with respect to the momentum sectors $\mathcal{H}_N^{(P)}$. For a general two-body interaction, this block structure is reflected in momentum conservation and can be expressed in second quantization as
\begin{equation}
  \hat W = \!\sum_{k_1,k_2,k_3,k_4} \! \!\!
  W_{k_1 k_2 k_3 k_4}\,
  \delta_{k_1+k_2,k_3+k_4}\,
  \hat b_{k_1}^\dagger \hat b_{k_2}^\dagger \hat b_{k_3}\hat b_{k_4},
\end{equation}
where the Kronecker delta is understood modulo $d$. A paradigmatic example is
the one-dimensional Bose-Hubbard model with periodic boundary conditions, for
which
\begin{equation}
  \label{eq:hubbard_W}
  \hat W = \frac{1}{d}\sum_{k_1,k_2,k_3,k_4}\!
  \delta_{k_1+k_2,k_3+k_4}\,
  \hat b_{k_1}^\dagger \hat b_{k_2}^\dagger \hat b_{k_3}\hat b_{k_4},
\end{equation}
corresponding in position space to an on-site interaction $\sum_{i=1}^d \hat n_i(\hat n_i-1)$.

The analysis and results presented in this section extend straightforwardly to
higher-dimensional lattices. For a three-dimensional periodic lattice, the
momentum label $k$ is replaced by a vector $\vec k=(k_x,k_y,k_z)$ with
$k_\alpha=0,\dots,d_\alpha-1$, and the symmetry sectors are labeled by the
conserved total momentum vector $\vec P$. All definitions and results carry
over directly under this replacement. Similarly, spinful bosons can be treated
by augmenting the momentum label $\vec k$ with a spin (or polarization)
quantum number $\sigma$, corresponding to a fixed quantization axis. In this
way, the framework presented here applies equally to systems with arbitrary spatial and
internal symmetries.

\section{Domain of the Functional and its Geometric Features}\label{sec:domain}

The goal of this section is to determine the domain of the symmetry-adapted
functional $\FpP$ in each symmetry sector $\HNP$, i.e., to characterize the set
of momentum occupation number vectors $\vec n$ that can arise from an
$N$-particle pure state $\ket{\Psi}$ with total momentum $P$. Understanding
this domain is essential, as it encodes purely kinematic constraints and will
later be shown to strongly shape the universal interaction functionals
$\FpP$.

The occupation numbers necessarily satisfy $\sum_k n_k = N$ and
$n_k \ge 0$, which define a $(d-1)$-simplex in $\mathbb{R}^d$. Once the total
momentum $P$ is fixed, however, these conditions are generally not sufficient.
For most choices of $(d,N,P)$, the actual domain $\dom \,\Fp^{(P)}$ is a strict
subset of this simplex and takes the form of a convex polytope that can be
visualized as arising from additional linear constraints near certain
vertices. While these constraints delimit the admissible region at the
kinematic level, we will show in Sec.~\ref{sec:BECforce} that, for concrete
systems, the universal interaction functional associated with a given
interaction $\hat W$ generates a \textit{diverging repulsive force} on the
ground-state occupation number vector $\vec n$ near the boundary of the
domain. This observation provides the central motivation for analyzing the
geometry of $\dom\,\Fp^{(P)}$ in detail.

In order to characterize the domain of $\Fp^{(P)}$ in a precise geometric way,
we recall that a \textit{convex combination} of vectors is a linear combination
with nonnegative coefficients summing to one, and that the \textit{convex hull}
of a collection of vectors is the set of all such convex combinations.
\begin{thm}
  \label{thm:domain_of_F}
  Consider a one-dimensional system of $N$ bosons on $d$ lattice sites. For
  each value of the total momentum $P=0, \dotsb, d-1$, the domain of
  $\Fp^{(P)}$ is
  \begin{equation}
    \label{eq:convex_hull}
    \mathrm{conv}\left\{
      \vec n \in \mathbb{N}_0^d \;\Big\vert\; \sum_{k=0}^{d-1} n_k = N,\;
      \sum_{k=0}^{d-1} k n_k \equiv P \pmod d
      \right\},
  \end{equation}
  i.e., the convex hull of the occupation number vectors $\vec n$ of all
  configuration states $\ket{\vec n}$~\eqref{eq:bosonic_fock_state} with total
  momentum $P$.
  \begin{proof}
      For any $N$-particle state $\ket{\Psi}$ with momentum $P$, i.e.,
      $\ket{\Psi}\in \HNP$, we have $\ket{\Psi} = \sum_{\alpha=1}^{\dim \HNP}
      c_\alpha \ket{\nalpha}$, where each $\nalpha$ satisfies
      $\sum_{k=0}^{d-1} k n^{(\alpha)}_k \equiv P \pmod d$. Thus the momentum
      occupation number vector $\vec n$ of $\ket{\Psi}$ is given by
      \begin{equation}
        \label{eq:n_vector_from_state_coefficients}
        \begin{aligned}
          n_k &= \braket{\Psi|\hat b_k^\dagger \hat b_k|\Psi}
        = \sum_{\alpha,\beta} c_\alpha^* c_\beta
          \braket{\nalpha|\hat b_k^\dagger \hat b_k|\nbeta} \\
          &= \sum_\alpha |c_\alpha|^2 n^{(\alpha)}_k,
        \end{aligned}
      \end{equation}
      where we have used
      $\braket{\nalpha|\hat b_k^\dagger \hat b_k|\nbeta}
      = n^{(\alpha)}_k \delta_{\alpha\beta}$. Inspecting
      Eq.~\eqref{eq:n_vector_from_state_coefficients}, we immediately see that
      the occupation number vector $\vec n$ sweeps through the entire convex
      hull of $\{\nalpha\}$ as the wave-function coefficients $\{c_\alpha\}$
      are varied.
  \end{proof}
\end{thm}

In words, Theorem~\ref{thm:domain_of_F} shows that the domain of the functional
is a convex polytope generated as convex combinations by the occupation number vectors of all
configuration-number states with fixed total momentum. The statement extends
straightforwardly to spatial dimensions larger than one, where the domain is
given by the convex hull of all configuration-number states $\ket{\vec n}$ with
fixed total momentum vector $\vec P$. Moreover, since the domain is convex and
the map $\ket{\Psi}\!\bra{\Psi} \mapsto \vec n$ is linear, the same polytope is
obtained if one considers all ensemble states with support on
$\mathcal{H}_N^{P}$ rather than just the pure states $\ket{\Psi}\!\bra{\Psi}$.

\begin{figure}[htb]
  \subfloat[$d=2,N=4,P=0$\label{fig:d2N4P0}]{
    \includegraphics[width=.49\columnwidth]{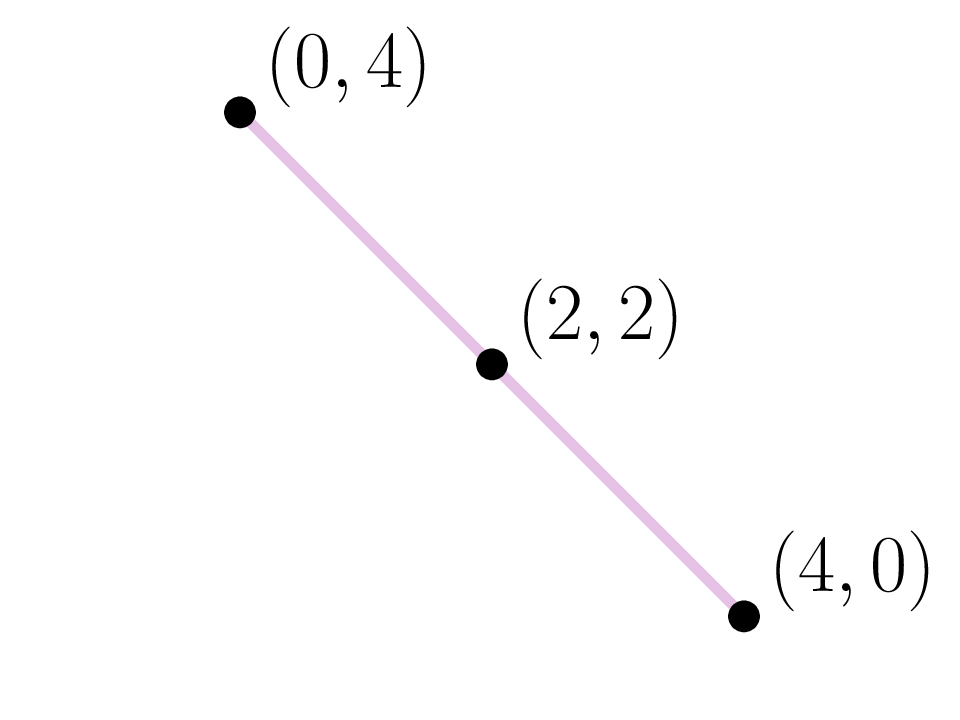}
  }
  \subfloat[$d=2,N=4,P=1$\label{fig:d2N4P1}]{
    \includegraphics[width=.49\columnwidth]{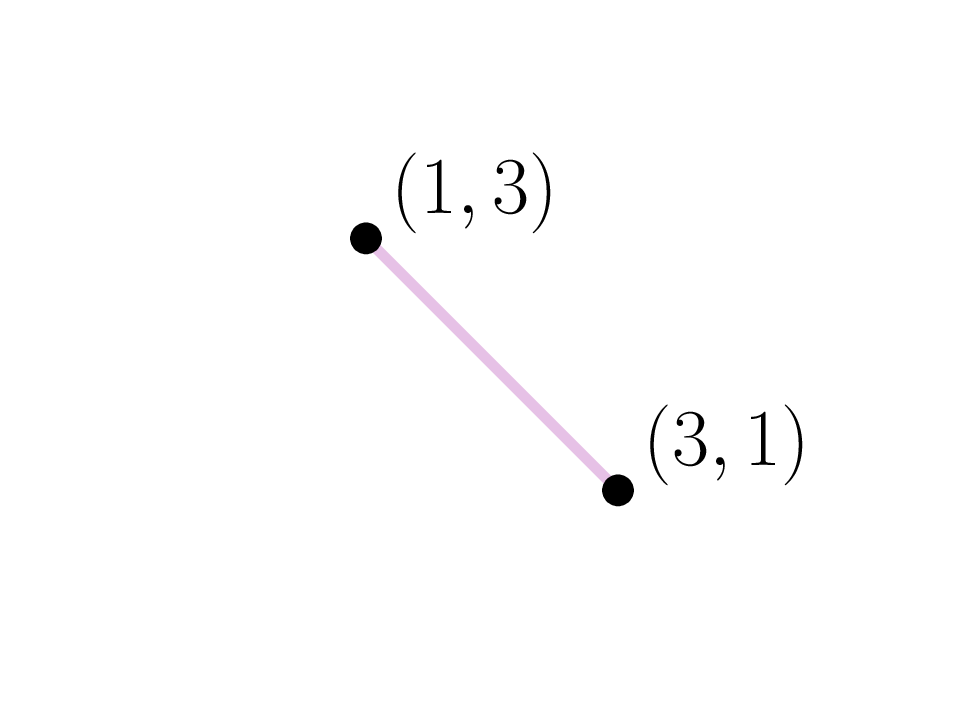}
  }

  \subfloat[$d=3,N=3,P=0$\label{fig:d3N3P0}]{
    \includegraphics[width=.49\columnwidth]{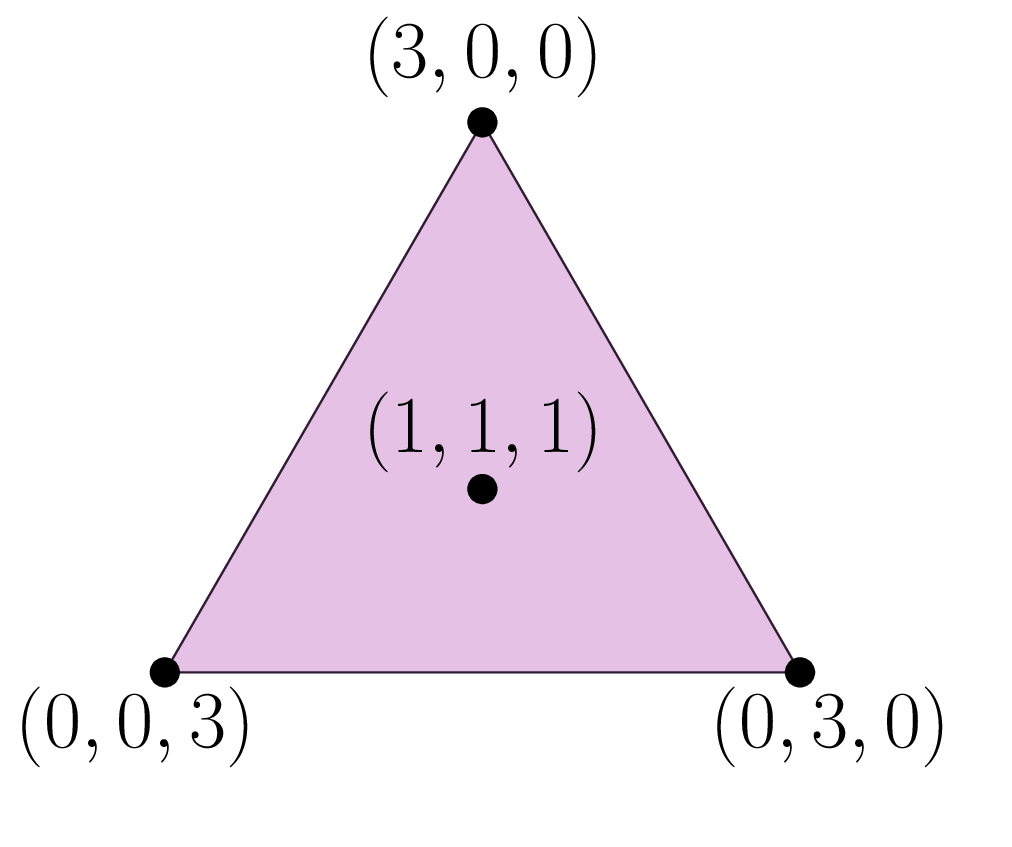}
  }
  \subfloat[$d=3,N=3,P=1$\label{fig:d3N3P1}]{
    \includegraphics[width=.49\columnwidth]{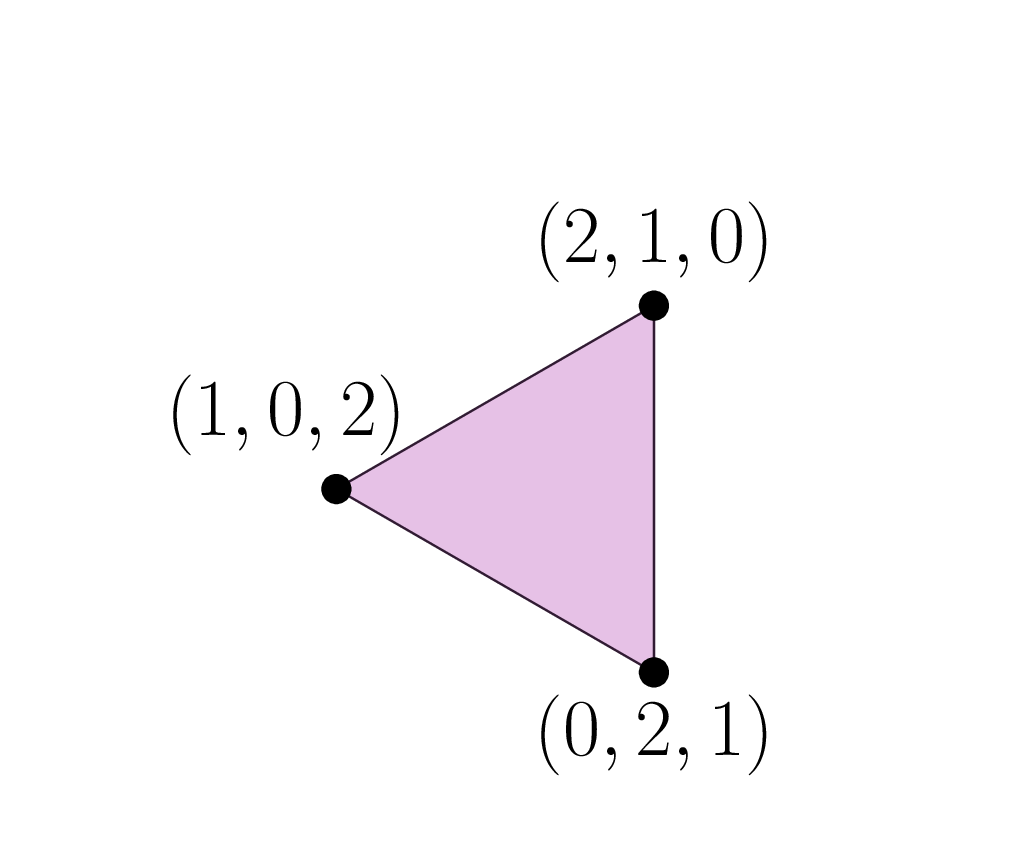}
  }

  \subfloat[$d=3,N=12,P=0$\label{fig:d3N12P0}]{
    \includegraphics[width=.5\columnwidth]{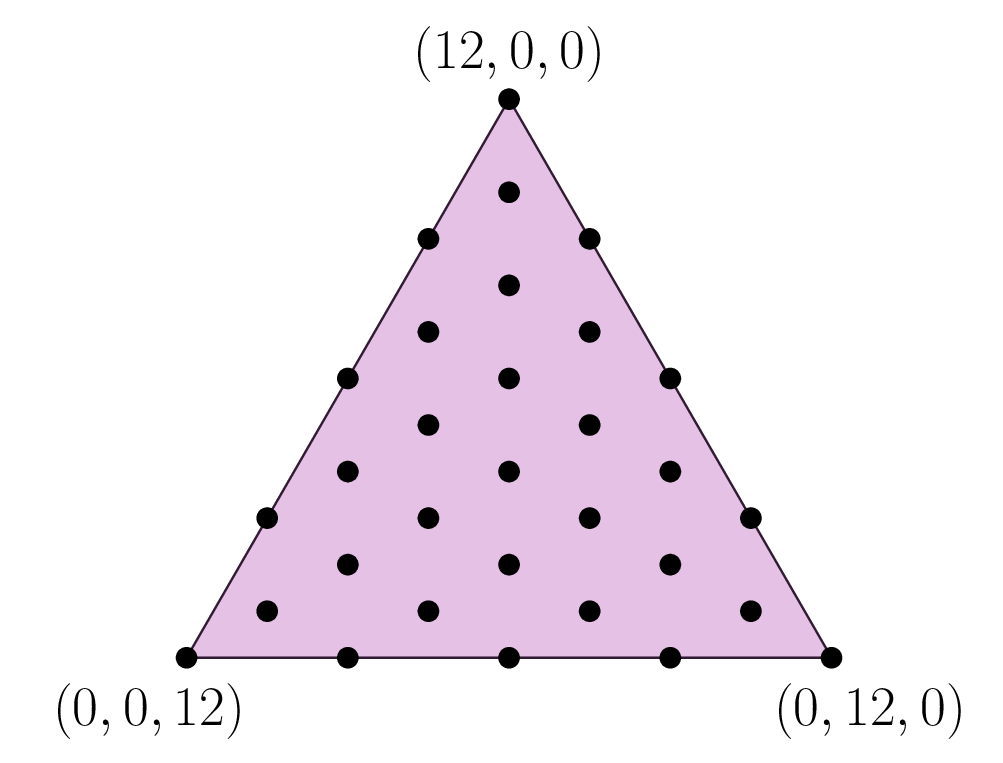}
  }
  \subfloat[$d=3,N=12,P=1$\label{fig:d3N12P1}]{
    \includegraphics[width=.5\columnwidth]{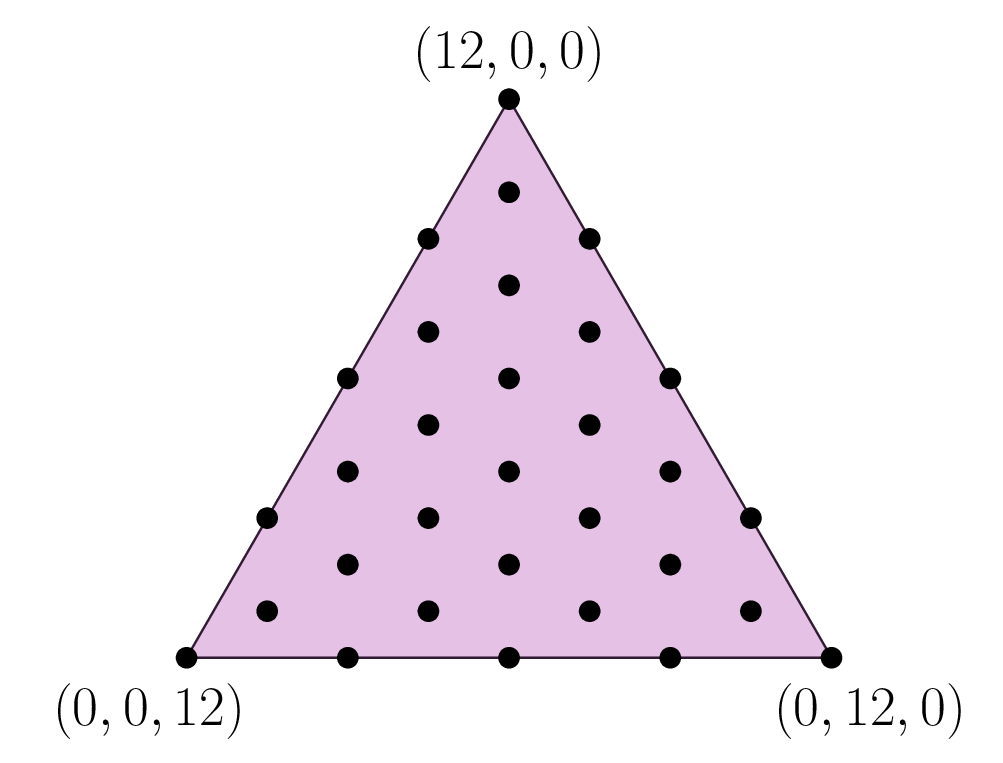}
  }

  \subfloat[$d=4,N=4,P=0$\label{fig:d4N4P0}]{
    \includegraphics[width=.5\columnwidth]{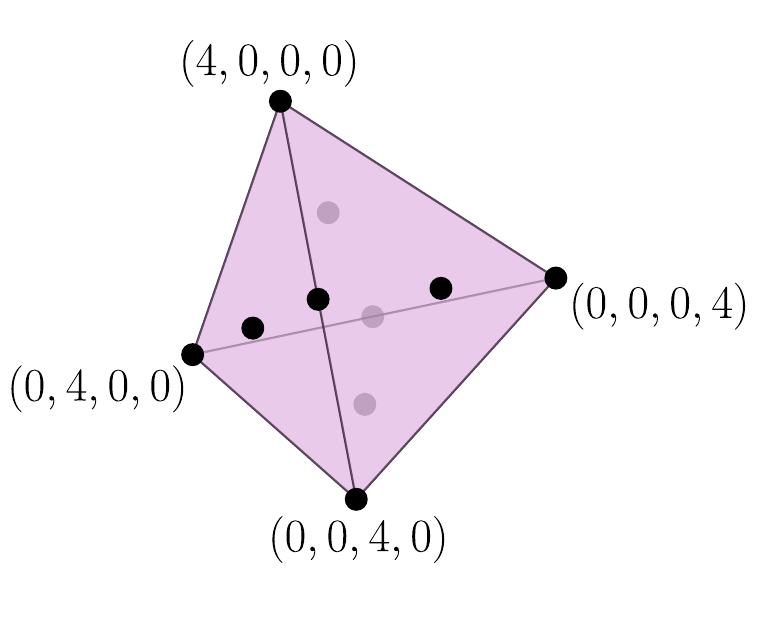}
  }
  \subfloat[$d=4,N=4,P=1$\label{fig:d4N4P1}]{
    \includegraphics[width=.5\columnwidth]{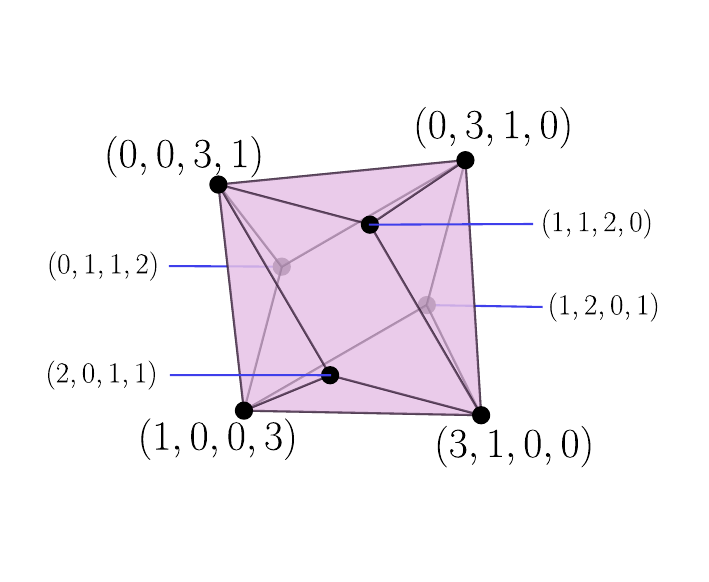}
  }

  \caption{Structure of $N$-particle momentum space configurations and domains
  of the functional.}
  \label{fig:domains_dNP}
\end{figure}

To illustrate these general results, we first consider the case of $d=2$
lattice sites. The $N$-particle Hilbert space is spanned by the momentum
eigenstates $\ket{N,0}, \ket{N-1,1}, \dotsb, \ket{0, N}$, and there are two
possible values of the total momentum, $P=0$ and $P=1$. Assuming $N$ is even,
the $P=0$ subspace is spanned by $\ket{N,0}, \ket{N-2,2}, \dotsb, \ket{0,N}$,
while the $P=1$ subspace is spanned by $\ket{N-1,1}, \ket{N-3,3},
\dotsb, \ket{1,N-1}$. According to Theorem~\ref{thm:domain_of_F}, this yields
\begin{equation}
  \begin{aligned}
    &\mathrm{dom}(\Fp^{(0)}) = \{(n_0, n_1)\in \mathbb{R}^2 \mid n_0+n_1 = N,\;
    0 \le n_1 \le N\},\\
    &\mathrm{dom}(\Fp^{(1)}) = \{(n_0, n_1)\in \mathbb{R}^2 \mid n_0+n_1 = N,\;
    1 \le n_1 \le N-1\}.
  \end{aligned}
\end{equation}
The corresponding domains for $N=4$ are shown in panels (a) and (b) of
Fig.~\ref{fig:domains_dNP}.

For general $(d,N,P)$, the geometry of the domain becomes more involved, as
illustrated in Fig.~\ref{fig:domains_dNP}. A few general observations are in
order. First, whenever $d$ divides $N$, the configurations
$\ket{N,0,\dotsc,0}, \ket{0,N,0,\dotsc,0}, \dotsc$ all belong to the
zero-momentum subspace, and $\dom \,\Fp^{(0)}$ coincides with the full simplex
spanned by these states (see panels (a), (c), (e), and (g)). Second, in the
limit $N \to \infty$ at fixed $d$ and $P$, the vertices of the simplex approach,
relative to $N$, the nearest admissible occupation number vectors with total
momentum $P$, so that the domain approaches a simplex asymptotically (see panel
(f)).

Finally, panels (a), (c), (e), and (f) of Fig.~\ref{fig:domains_dNP} highlight
a distinctive feature of the bosonic setting in contrast to fermions
\cite{schillingDivergingExchangeForce2019}: the occupation number vectors of
configuration states $|\vec{n}\rangle \in \mathcal{H}_N^{(P)}$ are not
necessarily extremal points of the domain but form a regular lattice that may
also populate its interior. As will become clear in Sec.~\ref{sec:BECforce},
both the shape of the domain and the distribution of occupation number vectors
of the configuration states within it play an essential role in determining the
behavior of the functional near the boundary.

\section{Form of universal functional}\label{sec:F}

In usual RDMFT without explicitly incorporating symmetries, as outlined in
Sec.~\ref{sec:notation_general_summary}, various approximate universal
functionals have been proposed in the literature, mostly for electronic
systems
\cite{mullerExplicitApproximateRelation1984,marquesEmpiricalFunctionalsReduceddensitymatrixfunctional2008,pirisNaturalOrbitalFunctional2011,pirisGlobalNaturalOrbital2021,schmidtMachineLearningUniversal2021,sharmaReducedDensityMatrix2008},
with exact analytical results available only for a few specific model systems
\cite{cohenLandscapeExactEnergy2016,benavides-riverosReducedDensityMatrix2020}.
The systematic exploitation of spatial symmetries in RDMFT has so far been
limited. For arbitrary translationally invariant fermionic lattice models, this was
carried out in a comprehensive manner in Ref.~\cite{schillingDivergingExchangeForce2019},
where the pure functional $\Fp$ was shown to depend sensitively on the geometry
of its domain. The aim of this section is to extend these ideas to
\emph{bosonic} systems and to accommodate arbitrary kinetic-energy operators,
in line with modern ultracold-atom experiments, where both the amplitudes and
the range of hopping processes can be tuned \cite{Guenter2013,Schempp2015}.
From now on, to simplify the notation, we drop the superscript $P$ in
$\Fp^{(P)}$.

As is apparent in Sec.~\ref{sec:domain}, the domain of $\Fp$ is a subset of
the hyperplane $\{\vec n \in \mathbb{R}^n \mid \sum_k n_k = N\}$ bounded by a
list of $J = J(d, N, P)$ affine constraints
\begin{equation}
	\label{eq:Dj_constraints}
	D^{(j)}(\vec n) = \vec \kappa^{(j)} \cdot \vec n + \mu^{(j)} \ge 0,
\end{equation}
where $j=1, \dotsb, J$ labels the facets. Note that there is some ambiguity in
the choice of $\vec \kappa^{(j)}$ and $\mu^{(j)}$. For instance, scaling them by a positive constant,
or replacing them with $\vec \kappa^{(j)}\rightarrow \vec \kappa^{(j)} +
(\nu,\dotsb, \nu)$ and $\mu^{(j)}\rightarrow \mu^{(j)} - \nu N$ for any $\nu$,
results in an equivalent constraint. This ambiguity, however, does not affect
any physical result in this section.

As an illustrative example, consider $N=3$ bosons on $d=3$ lattice sites with total momentum
$P=1$ (panel (d) of Fig.~\ref{fig:domains_dNP}). The spanning states are
$\vec n^{(1)} = \ket{2,1,0}$, $\vec n^{(2)} = \ket{0,2,1}$, and
$\vec n^{(3)} = \ket{1,0,2}$. The domain of the functional is given by the three
inequality constraints
\begin{equation}\label{domain331}
  \begin{aligned}
  &D^{(1)}(\vec n) = n_0 - n_2 + 1 \ge 0 \\ 
  &D^{(2)}(\vec n) = n_1 - n_0 + 1 \ge 0 \\ 
  &D^{(3)}(\vec n) = n_2 - n_1 + 1 \ge 0,
  \end{aligned}
\end{equation}
together with the normalization condition $n_0+n_1+n_2=3$.

The key idea of our work is to express now $\Fp[\vec n]$ in terms of the quantities
$D^{(j)}(\vec n)$, $j=1, \dotsb, J$, which geometrically measure the distance
of $\vec n$ (up to a multiplicative factor) from the $j$-th facet. We proceed in
two steps. First, we treat a simplified scenario, the so-called `simplex
setting' (see below for a precise definition). Although a general combination
of $(d, N, P)$ does not satisfy the simplex assumption, this basic setting
allows us to derive the functional in a straightforward way and to illustrate
the essential features without encountering technical complications. Next, in
the subsequent section, armed with the intuition obtained in the simplex
setting, we will drop this assumption and derive a functional form in full
generality, i.e., for arbitrary $(d,N,P)$.

\subsection{The Simplex}
\label{sec:F_simplex}
In general, an $m$-\textit{simplex} is an $m$-dimensional polytope which is the
convex hull of its $m+1$ vertices. In this section, we assume that the domain
is a simplex with exactly $\dim \HN^{(P)}$ vertices, meaning in particular that
the momentum occupation number vector $\vec n^{(\alpha)}$ of each
configuration state $\ket{\vec n^{(\alpha)}}$ lies on a vertex. This is a
strong condition on $(d, N, P)$, and only specific combinations of low
dimension and low particle number result in such a scenario. For example,
panels~(b) and (d) of Fig.~\ref{fig:domains_dNP} satisfy this assumption, while
the remaining panels do not. Note that although the functional domains in
panels~(a), (c), (e), and (g) are geometrically simplices, they are excluded
from the discussion in this section because of the additional $\nalpha$
vectors which are not vertices. Hereafter, \textit{the simplex assumption} or
\textit{simplex setting} always refers to this stronger condition, rather than
merely requiring the domain to be a simplex.

Let $\vec n^{(\alpha)}$, $\alpha=1, \dotsb, \dim \HNP$, denote the list of
vertices. The Hilbert space $\HN^{(P)}$ is then spanned by the corresponding configuration states
$\ket{\vec n^{(\alpha)}}$. Since for simplices there is a one-to-one
correspondence between vertices and facets, the number $J$ of facets equals
$\dim \HNP$, and we use Greek letters $\alpha,\beta$ to label the constraints
instead of $j$, as in Eq.~\eqref{eq:Dj_constraints}. More precisely, we label
the constraints such that $D^{(\alpha)}=0$ corresponds to the facet opposite
the vertex $\vec n^{(\alpha)}$.

To evaluate $\Fp$ at a given occupation number vector $\vec n$, we consider a
state $\ket{\Psi} = \sum_\alpha c_\alpha \ket{\vec n^{(\alpha)}}$ whose
one-body occupations satisfy
$\braket{\Psi|\hat b_k^\dagger \hat b_k|\Psi} = n_k$ for all $k$. Applying the affine
functional $D^{(\beta)}$ then to both sides of Eq.~\eqref{eq:n_vector_from_state_coefficients} yields
\begin{equation}
  \label{eq:D_beta_linear_relation}
  \sum_\alpha |c_\alpha|^2 D^{(\beta)}\!\left(\vec n^{(\alpha)}\right)
  = D^{(\beta)}(\vec n).
\end{equation}
Since all vertices except $\vec n^{(\beta)}$ lie on the facet
$D^{(\beta)}=0$, we have
$D^{(\beta)}(\vec n^{(\alpha)}) = \VFactor{\alpha}\delta_{\alpha\beta}$ with
$\VFactor{\alpha} \equiv D^{(\alpha)}(\vec n^{(\alpha)})$. It therefore
follows that
\begin{equation}
  \label{eq:simplex_coeff_D_relation}
  |c_\beta|^2 =
  \frac{D^{(\beta)}(\vec n)}{L_\beta},
\end{equation}
which admits a direct geometric interpretation as a ratio of distances to the
facets of the domain.

As a central result of our work, the interaction functional \eqref{eq:symmetry_sector_functional_def} in the
simplex setting can thus be written explicitly as
\begin{equation}
  \label{eq:F_simplex}
  \Fp[\vec n] = \sum_{\alpha,\beta=1}^{\dim \HN^{(P)}} W_{\alpha\beta}
  \eta^*_\alpha(\vec n)\, \eta_\beta(\vec n)
  \sqrt{\frac{D^{(\alpha)}(\vec n)}{\VFactor{\alpha}}}
  \sqrt{\frac{D^{(\beta)}(\vec n)}{\VFactor{\beta}}},
\end{equation}
where $W_{\alpha\beta}\equiv \braket{\vec n^{(\alpha)}|\hat W|\vec
n^{(\beta)}}$. The phase factors $\{\eta_\alpha(\vec n)\}$ arise from the
minimization of $\Fp$ and depend implicitly on $\vec n$; for specific
interactions $\hat W$, they may be determined explicitly.

The key result~\eqref{eq:F_simplex} makes transparent how the geometry of the
domain directly shapes the structure of the universal interaction functional,
to a large extent independently of the specific interaction $\hat W$. The
dependence on the occupation numbers enters exclusively through the facet
distances $D^{(\alpha)}(\vec n)$, so that the value of $\Fp$ is governed by the
relative position of $\vec n$ within the domain. In particular, the
square-root singularities associated with the facets already signal the
pronounced influence of the boundary geometry on the behavior of the
functional, a feature that will remain central beyond the simplex setting.

\begin{figure}
  \centering
	\centering
  \subfloat[$w=-1$]{
    \includegraphics[height=.47\columnwidth]{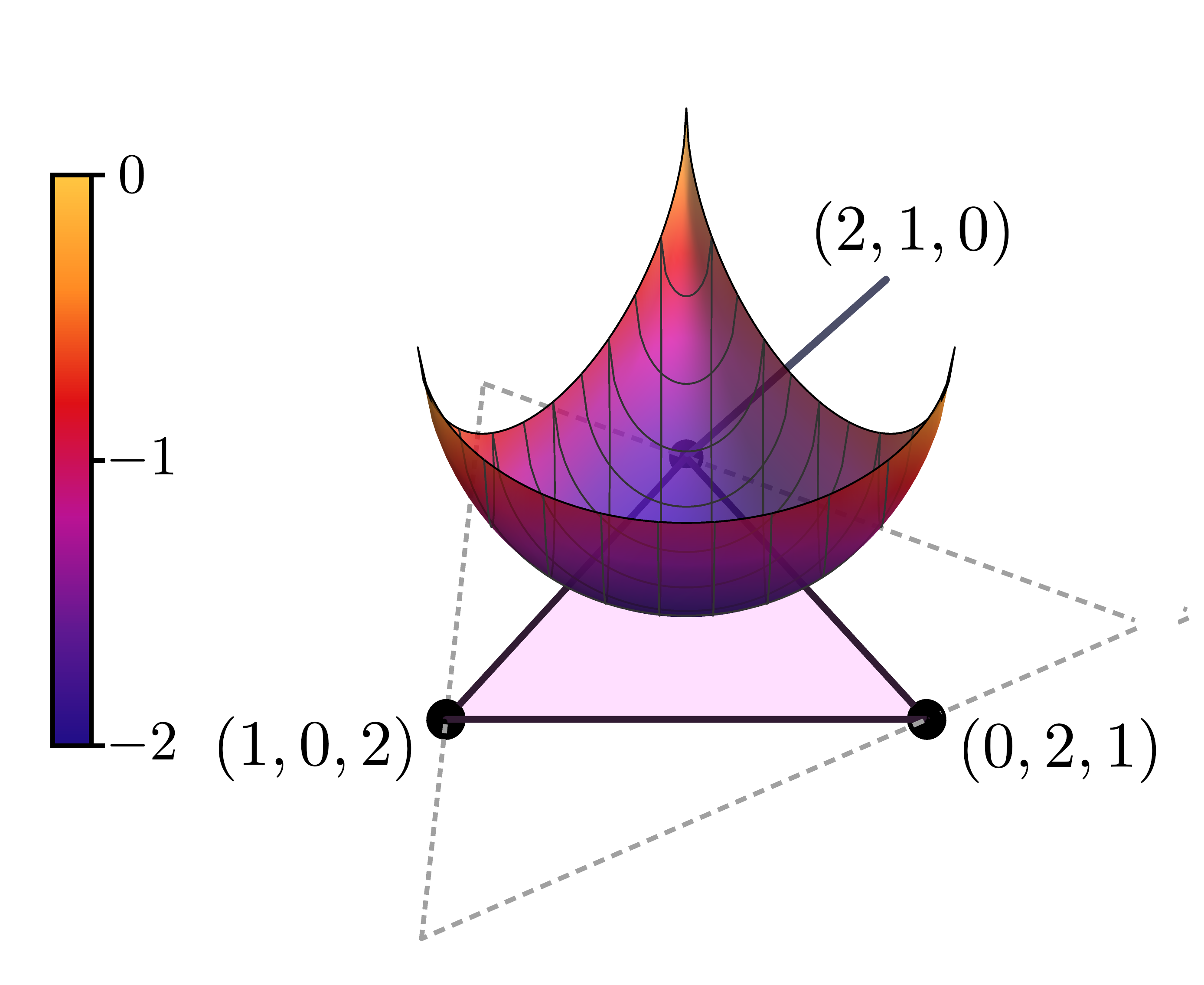}
  }
  \subfloat[$w=1$]{
  	\includegraphics[height=.484\columnwidth]{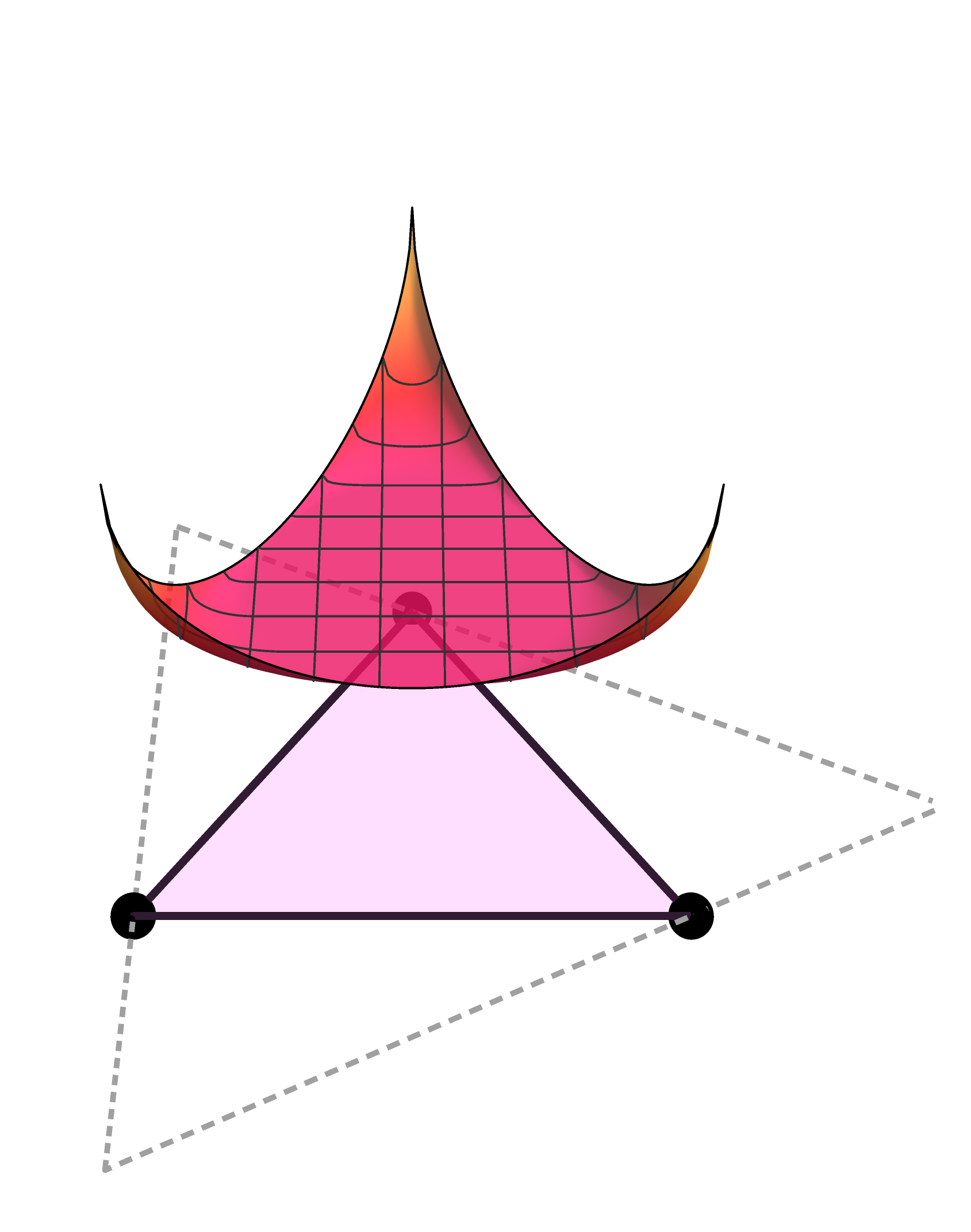}
  }
  \caption{The universal functional $\mathcal{F}[\vec n]$ for $(d,N,P)=(3,3,1)$ with
  negative (left) and positive (right) coupling parameter $w$. The solid triangle 
  is the domain of $\mathcal{F}$, while the dashed lines outline
  the convex hull of $(3,0,0), (0,3,0),$ and $(0,0,3)$. See text for details.
  }
  \label{fig:331_functional}

\end{figure}

As an example, we consider again the setting $(d,N,P)=(3,3,1)$, whose domain is characterized by the 
inequality constraints in \eqref{domain331}. We have $L_1=L_2=L_3 = 3$. Hence, Eq.~\eqref{eq:F_simplex} becomes
\begin{eqnarray}
    \Fp[\vec n] &=& u
    +\frac{2}{3}  \min_{\eta_1,\eta_2,\eta_3} w\,\mathrm{Re}\Bigg[
    \eta_1^*\eta_2 \sqrt{D^{(1)}(\vec n)D^{(2)}(\vec n)}\\
    && \!+ \eta_2^*\eta_3 \sqrt{D^{(2)}(\vec n)D^{(3)}(\vec n)}
     + \eta_3^*\eta_1 \sqrt{D^{(3)}(\vec n)D^{(1)}(\vec n)}
    \Bigg], \nonumber
\end{eqnarray}
where we have assumed $W_{11}=W_{22}=W_{33}=u$ and
$W_{\alpha\neq \beta} = w$, as in the Bose-Hubbard model. If $w\le 0$, the
minimization over the phases $\{\eta_\alpha\}$ is trivial and one finds
\begin{eqnarray}
    \Fp[\vec n] &=& u
    - \frac{2}{3} |w| \Bigg[
    \sqrt{D^{(1)}(\vec n)D^{(2)}(\vec n)}\\
    && \!+ \sqrt{D^{(2)}(\vec n)D^{(3)}(\vec n)}
     + \sqrt{D^{(3)}(\vec n)D^{(1)}(\vec n)}
    \Bigg]. \nonumber
\end{eqnarray}
For $w$ positive, the minimization over the phases $\{\eta_\alpha\}$ is still
relatively simple. A straightforward calculation shows that
$\Fp[\vec n]= u- w$ if
$\sqrt{D^{(1)}(\vec n)}, \sqrt{D^{(2)}(\vec n)}, \sqrt{D^{(3)}(\vec n)}$
satisfy the triangle inequality (i.e.,
$\sqrt{D^{(1)}(\vec n)}\le
\sqrt{D^{(2)}(\vec n)} + \sqrt{D^{(3)}(\vec n)}$ plus cyclic permutations).
Otherwise,
\begin{eqnarray}
    \Fp[\vec n] &=& u+ \frac{2}{3}w\Big(
    \sqrt{D^{(1)}(\vec n)D^{(2)}(\vec n)}\\
    &&-\sqrt{D^{(2)}(\vec n)D^{(3)}(\vec n)}
    - \sqrt{D^{(3)}(\vec n)D^{(1)}(\vec n)}\Big) \nonumber
\end{eqnarray}
if $D^{(3)}(\vec n)\ge D^{(1)}(\vec n), D^{(2)}(\vec n)$ (and analogously for
the other cases). The functionals for $w=1$
and $w=-1$ are shown in Fig.~\ref{fig:331_functional}.

\subsection{Beyond the simplex}
\label{sec:F_beyond_simplex}

Outside the simplex setting, i.e., whenever the domain is not a simplex with
$\dim \HNP$ vertices, the coefficients $\{|c_\alpha|^2\}$ can no longer be
reconstructed uniquely from the facet distances
$\{D^{(j)}(\vec n)\}$, in contrast to the situation in
Eq.~\eqref{eq:simplex_coeff_D_relation}. Instead, the quantities
$D^{(j)}(\vec n)$ depend \emph{linearly} on $\{|c_\alpha|^2\}$, but the
associated linear map has a nontrivial kernel. As a result, deriving an analog
of Eq.~\eqref{eq:F_simplex} requires a partial inversion of this map, a
strategy first employed in the fermionic context in
Ref.~\cite{schillingDivergingExchangeForce2019}. We now adapt this approach to
the bosonic setting and refine it.

To this end, we fix an occupation number vector $\vec n$ in the domain and let
$\ket{\Psi}=\sum_{\alpha} c_\alpha \ket{\nalpha}$ be an $N$-particle state in
$\HNP$ with momentum occupation numbers $\vec n$. Using
Eq.~\eqref{eq:n_vector_from_state_coefficients}, we obtain
\begin{equation}
  \label{eq:D_from_c_temp}
  \begin{aligned}
    D^{(j)}(\vec n)
    &= \vec \kappa^{(j)} \cdot \vec n + \mu^{(j)} \\
    &= \braket{\Psi|\sum_{k=0}^{d-1} \kappa^{(j)}_k \hat n_k + \mu^{(j)}|\Psi} \\
    &= \sum_{\alpha} |c_\alpha|^2
       \bigl[\vec \kappa^{(j)}\cdot \nalpha + \mu^{(j)}\bigr] \\
    &= \sum_{\alpha} |c_\alpha|^2 D^{(j)}(\nalpha),
  \end{aligned}
\end{equation}
where $\hat n_k\equiv \hat b_k^\dagger \hat b_k$. Introducing the vector
$\vec y$ with components $y_\alpha\equiv |c_\alpha|^2 \ge 0$ and the matrix
$T\in\mathbb{R}^{J\times \dim\HNP}$ defined by
$T_{j\alpha}\equiv D^{(j)}(\nalpha) \ge 0$, this relation can be written
compactly as
\begin{equation}
  \label{eq:D_from_y_temp}
  \vec D(\vec n) = T\,\vec y.
\end{equation}
At this point, $\vec n$ (and hence $\vec D(\vec n)$) is given, and the task is
to determine all nonnegative vectors $\vec y$ solving
Eq.~\eqref{eq:D_from_y_temp}.

A natural tool to address this problem is the \emph{Moore-Penrose
pseudoinverse}. For a real matrix $T$ with singular value decomposition
$T=U\Sigma V^\top$, the pseudoinverse is defined as
$T^+=V\Sigma^+U^\top$, where $\Sigma^+$ is obtained from $\Sigma$ by inverting the nonzero
diagonal elements (the singular values) of $\Sigma$ and transposing. The pseudoinverse provides a canonical
(least-norm) solution to underdetermined linear systems and, crucially, allows
one to separate the uniquely determined components from contributions lying
in the kernel of $T$.

Applying this construction to Eq.~\eqref{eq:D_from_y_temp}, the general
solution can be written as
\begin{equation}
  \label{eq:y_from_D_by_pseudoinverse}
  \vec y = T^+ \vec D(\vec n) + \vec x,
\end{equation}
where $\vec x$ is an arbitrary vector in the kernel of $T$ and must be chosen
such that the resulting $\vec y$ has nonnegative components. To verify that
Eq.~\eqref{eq:y_from_D_by_pseudoinverse} indeed solves
Eq.~\eqref{eq:D_from_y_temp}, note that by $N$-representability of $\vec n$
there exists a vector $\vec y\,'$ with $\vec D(\vec n)=T\vec y\,'$, and the
defining property $TT^+T=T$ of the Moore-Penrose pseudoinverse then implies
$T\vec y = \vec D(\vec n)$.

We may therefore write
\begin{equation}
  \ket{\Psi} =
  \sum_{\alpha=1}^{\dim \HNP}
  \eta_\alpha
  \sqrt{\sum_{j=1}^J T^+_{\alpha j}D^{(j)}(\vec n) + x_\alpha}
  \ket{\nalpha},
\end{equation}
where the phases $\{\eta_\alpha\}$ are at this stage free. Inserting this
expression into the constrained-search formulation
\eqref{eq:symmetry_sector_functional_def} yields
\begin{widetext}
\begin{equation}
  \label{eq:general_F_form}
  \Fp[\vec n] =
  \sum_{\alpha,\beta=1}^{\dim\HNP}
  W_{\alpha\beta}\eta_\alpha^*\eta_\beta
  \sqrt{\sum_{j=1}^J T^+_{\alpha j}D^{(j)}(\vec n) + x_\alpha}
  \sqrt{\sum_{j=1}^J T^+_{\beta j}D^{(j)}(\vec n) + x_\beta}.
\end{equation}
\end{widetext}
For each $\vec n$ in the domain, the phases $\eta_\alpha$ and the vector
$\vec x\in \mathrm{ker}(T)$ are chosen so as to minimize
Eq.~\eqref{eq:general_F_form}. If, in addition to translational invariance,
the Hamiltonian is invariant under parity-time ($\mathrm{PT}$) symmetry, the
phases can be taken real without changing the resulting ground-state energy,
yielding an equivalent functional~\cite{liebertRefiningRelatingFundamentals2023}.

Equation~\eqref{eq:general_F_form} reveals a deep connection between the
universal interaction functional $\Fp$ and the geometry of its domain. In this
representation, all nontrivial dependence on the occupation numbers enters
through the facet-distance coordinates $\vec D(\vec n)$, which are naturally
adapted to the shape of the admissible set. Apart from the minimizing choice
of the kernel vector $\vec x$, the general form of the functional is therefore
largely dictated by the geometry of the domain rather than by the specific
interaction $\hat W$. Viewed in this way, Eq.~\eqref{eq:general_F_form}
constitutes a geometry-driven reparametrization of the constrained-search
functional, suggesting a shift in how constrained-search approaches may be
formulated more broadly.

We now illustrate this general construction by means of an explicit example,
which also demonstrates how the formalism reproduces the simplex result as a
special case.

For $d=2$, $N=4$, and $P=0$, the domain of $\Fp$ is given by the convex hull of
$\vec n^{(1)}=(4,0)$, $\vec n^{(2)}=(2,2)$, and $\vec n^{(3)}=(0,4)$ (see panel
(a) of Fig.~\ref{fig:domains_dNP}). The domain is one-dimensional and has two
facets, corresponding to the vertices $\vec n^{(1)}$ and $\vec n^{(3)}$, which
give rise to the constraints $D^{(1)}(\vec n)=n_0 \geq 0$ and $D^{(2)}(\vec n)=n_1 \geq 0$.
The resulting matrix $T=(D^{(j)}(\vec n^{(\alpha)}))$ reads
\begin{equation}
  T =
  \begin{pmatrix}
    4 & 2 & 0 \\
    0 & 2 & 4
  \end{pmatrix}.
\end{equation}
Its singular value decomposition is
\begin{equation}
  T =
  \begin{pmatrix}
    -\frac{1}{\sqrt{2}} & \frac{1}{\sqrt{2}} \\
    \frac{1}{\sqrt{2}} & \frac{1}{\sqrt{2}}
  \end{pmatrix}
  \begin{pmatrix}
    4 & 0 & 0 \\
    0 & 2\sqrt{6} & 0
  \end{pmatrix}
  \begin{pmatrix}
    -\frac{1}{\sqrt{2}} & \frac{1}{\sqrt{3}} & -\frac{1}{\sqrt{6}} \\
    0 & \frac{1}{\sqrt{3}} & \frac{2}{\sqrt{6}} \\
    \frac{1}{\sqrt{2}} & \frac{1}{\sqrt{3}} & -\frac{1}{\sqrt{6}}
  \end{pmatrix}^\top,
\end{equation}
from which the pseudoinverse follows as
\begin{equation}
  \begin{aligned}
    T^+ &=
    \begin{pmatrix}
      -\frac{1}{\sqrt{2}} & \frac{1}{\sqrt{3}} & -\frac{1}{\sqrt{6}} \\
      0 & \frac{1}{\sqrt{3}} & \frac{2}{\sqrt{6}} \\
      \frac{1}{\sqrt{2}} & \frac{1}{\sqrt{3}} & -\frac{1}{\sqrt{6}}
    \end{pmatrix}
    \begin{pmatrix}
      \frac{1}{4} & 0 & 0 \\
      0 & \frac{1}{2\sqrt{6}} & 0
    \end{pmatrix}^\top
    \begin{pmatrix}
      -\frac{1}{\sqrt{2}} & \frac{1}{\sqrt{2}} \\
      \frac{1}{\sqrt{2}} & \frac{1}{\sqrt{2}}
    \end{pmatrix}^\top \\
    &= \frac{1}{24}
    \begin{pmatrix}
      5 & -1 \\
      2 & 2 \\
      -1 & 5
    \end{pmatrix}.
  \end{aligned}
\end{equation}
The kernel of $T$ is spanned by $(-1,2,-1)$, so that $\vec x=\xi(-1,2,-1)$.
The radicands in Eq.~\eqref{eq:general_F_form} are then given by
\begin{equation}
  \begin{aligned}
    T^+\vec D(\vec n)+\vec x
    &= \frac{1}{24}
    \begin{pmatrix}
      5 & -1 \\
      2 & 2 \\
      -1 & 5
    \end{pmatrix}
    \begin{pmatrix}
      n_0 \\ n_1
    \end{pmatrix}
    + \xi
    \begin{pmatrix}
      -1 \\ 2 \\ -1
    \end{pmatrix} \\
    &=
    \begin{pmatrix}
      \frac{1}{3} + \frac{1}{8}(n_1-n_2) - \xi \\
      \frac{1}{3} + 2\xi \\
      \frac{1}{3} - \frac{1}{8}(n_1-n_2) - \xi
    \end{pmatrix}.
  \end{aligned}
\end{equation}

Assuming $W_{13}=0$, $W_{11}=W_{33}$, and $W_{12}=W_{23}\equiv w$, as is the
case for the Bose-Hubbard model, Eq.~\eqref{eq:hubbard_W}, and setting
$W_{11}=W_{33}=0$ without loss of generality, a minimizing choice of phases is
$(\eta_1,\eta_2,\eta_3)=(1,-1,1)$. Equation~\eqref{eq:general_F_form} then
yields
\begin{eqnarray}
    \Fp[\vec n]
    &= & W_{22}\left(\frac{1}{3}+2\xi\right)
    -2|w|\sqrt{\frac{1}{3}+2\xi} \times\\
    &&\!
    \left[
      \sqrt{\frac{1}{3}+\frac{1}{8}(n_1-n_2)-\xi}
      +\sqrt{\frac{1}{3}-\frac{1}{8}(n_1-n_2)-\xi}
    \right].\nonumber
\end{eqnarray}

We emphasize that Eq.~\eqref{eq:general_F_form} provides an \emph{exact}
representation of the functional $\Fp$. While its explicit evaluation remains
challenging for general interactions $\hat W$, due to the dependence of
$\vec x$ and the phases $\eta_\alpha$ on $\vec n$ and the typically large
nullity of $T$, the formalism makes explicit how the geometry of the domain
constrains the functional form. In this way, it naturally suggests systematic
approximation strategies based on controlled estimates of the kernel
contribution $\vec x[\vec n]$, an idea we will pursue further in
Sec.~\ref{sec:ex}.

\section{Generalized BEC Force}\label{sec:BECforce}

A central theme of this work has been to elucidate how the geometry of the
domain of the universal functional $\Fp$ controls both its structure and its
physical implications. In the preceding sections, we established that the
functional admits an explicit representation in terms of distances $D^{(j)}$ of $\vec{n}$ to the
facets of its domain and that these distances encode nontrivial many-body
information. In this section, we demonstrate that this geometric perspective
has direct and quantitative consequences for the \emph{variational behavior}
of the functional: specifically, it gives rise to a repulsive force whenever
the natural occupation number vector $\vec n$ approaches the boundary of
$\dom(\Fp)$.

We begin by observing, based on Eq.~\eqref{eq:F_simplex}, that within the
simplex setting the functional takes the approximate form $\Fp[\vec n]\sim c_0 - c_1 \sqrt{D^{(\omega)}(\vec n)}$
whenever the distance $D^{(\omega)}(\vec n)$ to a facet approaches zero. As a
direct consequence, the gradient of the functional diverges according to
$\nabla_{\vec n}\Fp \sim 1/\sqrt{D^{(\omega)}(\vec n)}$ as
$D^{(\omega)}\rightarrow 0$. This already suggests that the boundary of the
domain is dynamically inaccessible for ground states, since the functional
develops an infinite slope that repels the ground-state occupation number
vector $\vec n$ back into the interior of the domain.

\begin{figure}
  \includegraphics[width=.90\columnwidth]{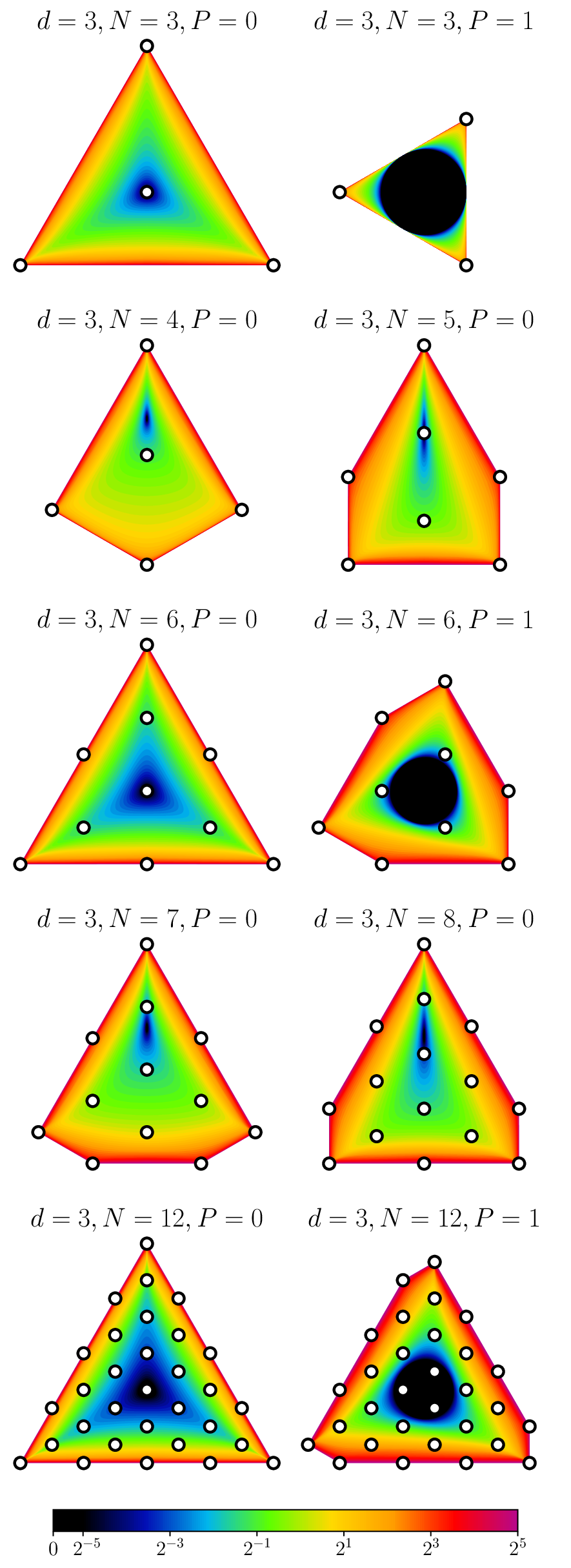}
  \caption{Magnitude of the gradient $|\nabla_{\vec n}\Fp|$ for $d=3$ and
  various combinations of $(N,P)$. See text for more details.}
  \label{fig:gradient_F_plots}
\end{figure}
To substantiate and generalize this observation, we now investigate the
behavior of the gradient $\nabla_{\vec n}\Fp$ numerically for the
one-dimensional Bose-Hubbard model with $d=3$, for which the domain of
admissible natural occupation numbers is two-dimensional. For several
particle numbers $N$ and total momenta $P$, we evaluate the magnitude
$|\nabla_{\vec n}\Fp|$ throughout the domain and visualize the result in
Fig.~\ref{fig:gradient_F_plots}. Our numerical procedure exploits the relation
\begin{equation}
\nabla_{\vec n}\Fp(\vec n) = -\vec t ,
\end{equation}
which holds whenever $\vec n$ corresponds to a ground state of the Hamiltonian $\hat{t}+\hat{W}$ for some kinetic
energy operator $\hat t = \sum_k t_k \hat n_k$. By scanning
over a sufficiently large set of kinetic energy vectors $\vec t$, we obtain
the exact value of the pure-state functional \eqref{eq:symmetry_sector_functional_def} for all
occupation number vectors $\vec n$ that arise from ground states (often
referred to as $v$-representable, or, in the present context,
$t$-representable). Points $\vec n$ that do not correspond to ground states of
any such $\hat t$ are shown in black in Fig.~\ref{fig:gradient_F_plots}. Since
these points typically do not lie close to the boundary of $\dom(\Fp)$, the
resulting data are fully sufficient for resolving the behavior of the
functional gradient in the boundary regime relevant for our purposes.

Inspection of Fig.~\ref{fig:gradient_F_plots} reveals a clear and robust
pattern: in all cases considered, the gradient of the functional diverges as
the boundary of the domain is approached. This divergence is neither confined
to particular facets nor to specific values of $(N,P)$, but instead appears
as a generic feature of the functional. 

Divergences of this type have recently been observed in a variety of reduced
density matrix functional settings. When the quantity playing the role of the
basic variable of the functional theory approaches the boundary of the
functional's domain, the corresponding derivative typically diverges as the
inverse square root of the distance to that boundary
\cite{schillingDivergingExchangeForce2019,
benavides-riverosReducedDensityMatrix2020,
liebertFunctionalTheoryBoseEinstein2021,
maciazekRepulsivelyDivergingGradient2021}.
In fermionic systems, this behavior, referred to as the \emph{exchange force},
has been demonstrated explicitly for translationally invariant one-band lattice
models as well as for the asymmetric lattice
dimer model with generic pair interaction \cite{schillingDivergingExchangeForce2019}.
For bosonic systems, the analogous divergence has so far been established
only in the vicinity of a specific boundary point of the domain, namely the
Bose-Einstein condensate vertex $(N,0,\dotsc)$, where it is known as the
\emph{BEC force} \cite{benavides-riverosReducedDensityMatrix2020, liebertFunctionalTheoryBoseEinstein2021,
maciazekRepulsivelyDivergingGradient2021}.

A key result of this section is that such a force is not confined to the
vicinity of the BEC vertex. Rather, it arises generically near \emph{any}
facet of the domain $\dom(\Fp)$. For this reason, we refer to it as the
\emph{generalized BEC force}. Its presence has important consequences, most
notably the prohibition of \emph{pinning}, i.e., situations in which the
ground-state occupation number vector lies exactly on the boundary of the set
of pure $(N,P)$-representable $\vec n$
\cite{maciazekImplicationsPinnedOccupation2020,
schillingImplicationsPinnedOccupation2020,
schillingHubbardModelPinning2015,
schillingReconstructingQuantumStates2017}.
This result is closely related to the question of whether natural occupation
numbers of the ground state can vanish, since any such vanishing would
necessarily place $\vec n$ on the boundary of the domain
\cite{cioslowskiNaturalOrbitalsTheir2024}.

In the following, we go beyond merely establishing the presence of a
diverging force at the boundary of the functional domain. We provide a direct
and general proof of this divergence and, for the first time, derive a simple
closed expression for its prefactor. This quantity determines the strength of
the generalized BEC force and constitutes a central result of this work. It
yields a sharp criterion for the presence or absence of pinning and
establishes an exact constraint that the functional of \emph{any}
translation-invariant bosonic system must satisfy, independently of the
interaction $\hat W$. More broadly, our derivation demonstrates that
repulsive boundary forces are a generic and structural feature of functional
theories.

Before proceeding, we fix a convention. As noted after
Eq.~\eqref{eq:Dj_constraints} in Sec.~\ref{sec:F}, the facet-defining
coefficients $\vec\kappa^{(j)}$ and $\mu^{(j)}$ are not unique. Throughout this
section, we choose their normalization such that $\vec\kappa^{(j)}$ is
tangent to the domain of $\Fp$ and satisfies
$\vec\kappa^{(j)}\!\cdot\!\vec\kappa^{(j)}=1$. In most cases, tangentiality is
equivalent to imposing $\sum_\alpha \kappa^{(j)}_\alpha = 0$.
For example,
the first constraint in Eq.~\eqref{domain331} for
$(d,N,P)=(3,3,1)$ becomes
\begin{equation}
D^{(1)}(\vec n) = \frac{n_0 - n_2 + 1}{\sqrt{2}},
\end{equation}
for which $\vec\kappa^{(1)} = \frac{1}{\sqrt{2}}(1,0,-1)$ and
$\mu^{(1)} = \frac{1}{\sqrt{2}}$.
This choice is purely conventional and serves to simplify the expressions
that follow.

\subsection{The Simplex}
\label{sec:BECforce_simlex}

As in Sec.~\ref{sec:F_simplex}, additional insight can be gained by first
analyzing the problem under the assumption that $\dom(\Fp)$ is a simplex with
$\dim\HNP$ vertices. While a heuristic argument for a diverging repulsive force
in the simplex setting was already given at the beginning of this section, we
now derive an explicit expression for the repulsion strength
$\ForcePrefactor$ (defined in Eq.~\eqref{eq:repulsion_strength_def}). Recall
(see Sec.~\ref{sec:F_simplex}) that the occupation number vectors
$\nalpha$, $\alpha=1,\dotsc,\dim\HNP$, are associated with constraints
$D^{(\alpha)}(\vec n)\ge 0$ corresponding to the facets opposite to $\nalpha$.
Pick an arbitrary facet $1\le \omega \le \dim\HNP$, i.e., the one opposite the
vertex $\vec n^{(\omega)}$, and consider the behavior of $\Fp$ in a small
neighborhood of a point $\vec n^*$ on this facet. By construction,
$D^{(\omega)}(\vec n^*)=0$, and we investigate
$\Fp[\vec n^*+\epsilon\vec v]$ for some inward-pointing vector $\vec v$. For
simplicity, we assume that $\vec n^*$ does not lie on any other facet, i.e.,
$D^{(\alpha)}(\vec n^*)>0$ for all $\alpha\neq\omega$.

In principle, the analysis could be carried out for an arbitrary inward-pointing
direction $\vec v$. However, the result would not differ qualitatively: if a
generalized BEC force exists, then displacements tangential to the facet induce
only subleading variations of $\Fp$. Accordingly, we choose
$\vec v=\vec\kappa^{(\omega)}$, where $\vec\kappa^{(\omega)}$ denotes the vector
of coefficients of the linear part of the constraint $D^{(\omega)}$, as defined
in Eq.~\eqref{eq:Dj_constraints}. By construction, $\vec\kappa^{(\omega)}$ is the
inward-pointing unit vector normal to the facet $\omega$.

We now employ Eq.~\eqref{eq:F_simplex} to analyze
$\Fp[\vec n^*+\epsilon\vec\kappa^{(\omega)}]$. More precisely, we compute the
derivative
\begin{equation}
  \label{eq:repulsion_strength_def}
  \ForcePrefactor[\vec n^*] \equiv
  \frac{d}{d\sqrt{\epsilon}}\Big|_{\epsilon=0}
  \Fp[\vec n^*+\epsilon\vec\kappa^{(\omega)}],
\end{equation}
which we refer to as the \textit{repulsion strength} of the generalized BEC force
at $\vec n^*$. If one were to compute $d\Fp/d\epsilon$ at
$\vec n^*+\epsilon\vec\kappa^{(\omega)}$ instead, the leading contribution would
scale as $\frac{1}{2\sqrt{\epsilon}}\ForcePrefactor[\vec n^*]$, up to terms that
are subleading in the limit $\epsilon\to0$. Hence, $\ForcePrefactor[\vec n^*]$
corresponds, up to a factor of $2$, to the prefactor of the diverging
$1/\sqrt{\epsilon}$ repulsive force.

A straightforward calculation yields
\begin{equation}
  \label{eq:D_from_epsilon}
  D^{(\alpha)}(\vec n^*+\epsilon\vec\kappa^{(\omega)})
  = D^{(\alpha)}(\vec n^*)
  + \epsilon\,\vec\kappa^{(\alpha)}\!\cdot\vec\kappa^{(\omega)},
\end{equation}
which for $\alpha=\omega$ reduces to
\begin{equation}
  \label{eq:D_omega_is_epsilon}
  D^{(\omega)}(\vec n^*+\epsilon\vec\kappa^{(\omega)})=\epsilon.
\end{equation}
While Eq.~\eqref{eq:D_omega_is_epsilon} is expected—$\epsilon$ parameterizes the
distance to $\vec n^*$, and $D^{(\omega)}$ measures the distance to the
facet—it also confirms that $\vec\kappa^{(\omega)}$ indeed points inward, as
assumed. Substituting Eq.~\eqref{eq:D_omega_is_epsilon} into
Eq.~\eqref{eq:F_simplex}, separating the terms involving $\omega$, and
performing the minimization over the phase $\eta_\omega$, we obtain
\begin{widetext}
\begin{equation}
  \label{eq:complicated_mess}
  \begin{aligned}
    \Fp[\vec n^*+\epsilon\vec\kappa^{(\omega)}]
    &= \min_{\{\eta_{\alpha\neq\omega}\}}
    \Bigg[
      \sum_{\substack{\alpha\neq\omega\\\beta\neq\omega}}
      \eta_\alpha^*\eta_\beta W_{\alpha\beta}
      \sqrt{\frac{D^{(\alpha)}(\vec n^*+\epsilon\vec\kappa^{(\omega)})}{L_\alpha}}
      \sqrt{\frac{D^{(\beta)}(\vec n^*+\epsilon\vec\kappa^{(\omega)})}{L_\beta}}
      \\
      &\hspace{4em}
      -2\sqrt{\frac{\epsilon}{L_\omega}}
      \left|
        \sum_{\alpha\neq\omega}
        \eta_\alpha W_{\omega\alpha}
        \sqrt{\frac{D^{(\alpha)}(\vec n^*+\epsilon\vec\kappa^{(\omega)})}{L_\alpha}}
      \right|
      + \frac{\epsilon}{L_\omega}W_{\omega\omega}
    \Bigg].
  \end{aligned}
\end{equation}
\end{widetext}

Our objective is to evaluate the derivative with respect to $\sqrt{\epsilon}$ at
$\epsilon=0$ (Eq.~\eqref{eq:repulsion_strength_def}). This can be achieved using
Danskin’s theorem~\cite{danskinTheoryMaxminApplications1966}, which allows the
interchange of minimization and differentiation for directional
derivatives. This is sufficient for the present analysis, even though it does
not necessarily imply differentiability of the functional $\Fp$.
The minimization must, however, be restricted to the set of minimizers of the
original problem. One thus finds
\begin{equation}
  \label{eq:BECforce_simplex}
  \ForcePrefactor[\vec n^*]
  = -\frac{2}{\sqrt{L_\omega}}
  \mathop{\widetilde{\min}}_{\{\eta_{\alpha\neq\omega}\}}
  \left|
    \sum_{\alpha\neq\omega}
    \eta_\alpha W_{\omega\alpha}
    \sqrt{\frac{D^{(\alpha)}(\vec n^*)}{L_\alpha}}
  \right|,
\end{equation}
where $\widetilde{\min}$ indicates that the phases
$\{\eta_{\alpha\neq\omega}\}$ are restricted to those minimizing
Eq.~\eqref{eq:complicated_mess}. Introducing the state
$\ket{\Phi^*(\vec n^*)}
= \sum_{\alpha\neq\omega}
\eta_\alpha\sqrt{D^{(\alpha)}(\vec n^*)/L_\alpha}\ket{\vec n^{(\alpha)}}$,
which is precisely the wave function minimizing the constrained search
\eqref{eq:symmetry_sector_functional_def} at $\vec n^*$, we may write
\begin{equation}
  \ForcePrefactor[\vec n^*]
  = -2\frac{\left|
    \braket{\vec n^{(\omega)}|\hat W|\Phi^*(\vec n^*)}
  \right|}{\sqrt{L_\omega}}.
\end{equation}

The minus sign reflects the repulsive nature of the force since
$\ForcePrefactor=d\Fp/d\sqrt{\epsilon}$ and increasing $\epsilon$ corresponds to
moving away from the facet. The geometric factor $\sqrt{L_\omega}$ in the
denominator implies that the repulsion weakens as the distance between the
vertex $\vec n^{(\omega)}$ and the facet increases. Physically, this expresses
that the state $\ket{\vec n^{(\omega)}}$ becomes less relevant for describing
densities near the facet, since any finite overlap with this state requires
deviations from the facet that scale with
$L_\omega=D^{(\omega)}(\vec n^{(\omega)})$. This also motivates our normalization
$|\vec\kappa^{(\omega)}|^2=1$, ensuring that $D^{(\omega)}(\vec n)$ coincides
with the actual distance to the facet.

\subsection{Beyond the simplex}

\begin{figure}
  \includegraphics[width=1.02\columnwidth]{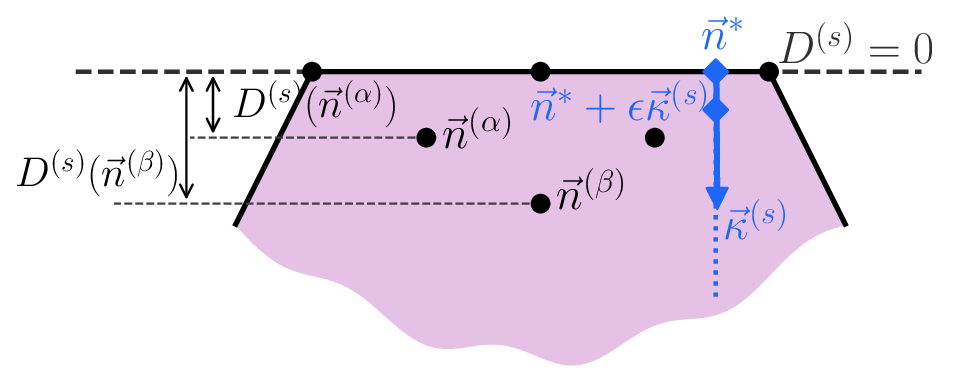}
  \caption{Geometric setup underlying the derivation of the generalized BEC
  force. Shown are a facet $D^{(s)}(\vec n)=0$, its inward normal
  $\vec\kappa^{(s)}$, and the normal path
  $\vec n^*+\epsilon\vec\kappa^{(s)}$ used in the analysis. See the main text for
  details.}
  \label{fig:facet_diagram}
\end{figure}

We now drop the simplex assumption and derive a general expression for the
repulsion strength $\ForcePrefactor$ for arbitrary settings $(d,N,P)$. As is
often the case, abandoning the simplex structure introduces substantial
complications. Most importantly, there is no longer a one-to-one correspondence
between facets of $\dom(\Fp)$ and occupation number vectors
$\{\nalpha\}$. Instead, the domain is characterized by $J$ constraints
$D^{(j)}\geq 0$, $j=1,\dotsc,J$, where $J$ is now smaller than the number of configuration states 
$\dim\HNP$. To avoid confusion, we therefore use Latin indices
$j,s,\dotsc$ to label facets and Greek indices $\alpha,\omega,\dotsc$ to label
configuration states.

Proceeding analogously to the simplex case, and as it is illustrated in Fig.~\ref{fig:facet_diagram}, we fix a facet
$1\le s\le J$ and a point $\vec n^*$ on it, such that $D^{(s)}(\vec n^*)=0$.
We assume that $\vec n^*$ does not lie on any other facet, i.e.,
$D^{(j)}(\vec n^*)>0$ for all $j\neq s$. Mimicking the definition of the
repulsion strength in the simplex setting [Eq.~\eqref{eq:repulsion_strength_def}],
we define 
\begin{equation}
  \label{eq:repulsion_strength_def_general}
  \ForcePrefactor[\vec n^*]
  \equiv
  \frac{d}{d\sqrt{\epsilon}}\Big|_{\epsilon=0}
  \Fp[\vec n^*+\epsilon\vec\kappa^{(s)}].
\end{equation}

After a sequence of rather technical and involved steps, which are presented in
Appendix~\ref{app:BECforce}, one arrives at
\begin{equation}
  \label{eq:amazing_exforce_formula}
  \ForcePrefactor[\vec n^*]
  =
  -2\left[
    \sum_{\alpha:\,D^{(s)}(\nalpha)>0}
    \frac{
      \left|
        \braket{\nalpha|\hat W|\Phi^*(\vec n^*)}
      \right|^2
    }{
      D^{(s)}(\nalpha)
    }
  \right]^{\!1/2},
\end{equation}
where $\ket{\Phi^*(\vec n^*)}$ denotes a minimizer of the constrained search
[Eq.~\eqref{eq:symmetry_sector_functional_def}] at $\vec n^*$. If multiple
minimizers exist, $\ket{\Phi^*(\vec n^*)}$ is chosen such that the expression in
Eq.~\eqref{eq:amazing_exforce_formula} is minimized.

Equation~\eqref{eq:amazing_exforce_formula} constitutes the central result of
this section. Together with the definition
\eqref{eq:repulsion_strength_def_general}, it implies that the gradient
$\nabla_{\vec n}\Fp$ diverges as $1/\sqrt{\epsilon}$ as $\vec n$ approaches a
facet at distance $\epsilon\to0$, provided that the sum in
Eq.~\eqref{eq:amazing_exforce_formula} does not vanish. Several notable features
of this result deserve emphasis. First, the overall minus sign reflects the
repulsive nature of the force, in agreement with the simplex analysis in
Sec.~\ref{sec:BECforce_simlex}. Second, each configuration state
$\ket{\nalpha}$ contributes with a geometric weight
$D^{(s)}(\nalpha)^{-1}$ (see Fig.~\ref{fig:facet_diagram}), i.e., inversely
proportional to its distance from the facet. Consequently, configuration states
far from the facet have a suppressed influence on the repulsion strength.
Third, since the radicand is a sum of nonnegative terms, the repulsion strength
can vanish only if
$\braket{\nalpha|\hat W|\Phi^*(\vec n^*)}=0$ for all $\alpha$ satisfying
$D^{(s)}(\nalpha)>0$.

Finally, it is instructive to recover the simplex result from
Eq.~\eqref{eq:amazing_exforce_formula}. In the simplex setting, there is exactly
one configuration state $\ket{\nalpha}$ not lying on the facet, so the sum
reduces to a single term and
Eq.~\eqref{eq:amazing_exforce_formula} reproduces
Eq.~\eqref{eq:BECforce_simplex}. The explicit appearance of the factor
$D^{(s)}(\nalpha)^{-1}$ thus confirms, in full generality, the geometric
intuition developed in the simplex case.

We now illustrate Eq.~\eqref{eq:amazing_exforce_formula} by considering the
one-dimensional Bose-Hubbard model with $(d,N,P)=(3,N,0)$, where $N$ is a
multiple of $3$. The cases $N=6$ and $N=12$ are shown in the third and fifth rows
of the left column of Fig.~\ref{fig:gradient_F_plots}, respectively. Although
the domain is a $2$-simplex (a triangle), this situation does not fall into what
we termed the `simplex setting', since additional occupation number vectors
exist beyond the vertices $(N,0,0)$, $(0,N,0)$, and $(0,0,N)$.

We choose $s$ to label the facet corresponding to the lower edge of the simplex,
for which
\begin{equation}
  D^{(s)}(\vec n)
  =
  \frac{1}{\sqrt{6}}(2n_0-n_1-n_2)
  +\frac{N}{\sqrt{6}},
\end{equation}
so that $\vec\kappa^{(s)}=\frac{1}{\sqrt{6}}(2,-1,-1)$ and
$\mu^{(s)}=N/\sqrt{6}$. Points on this facet may be parameterized as
$\vec n^*=(0,N\lambda,N(1-\lambda))$ with $\lambda\in(0,1)$, for which
$D^{(s)}(\vec n^*)=0$ holds identically.

To apply Eq.~\eqref{eq:amazing_exforce_formula}, we first note that the
configuration states lying on the facet are
\begin{equation}
  \ket{0,N,0},\;\ket{0,N-3,3},\;\dotsc,\;\ket{0,0,N}.
\end{equation}
The minimizers of the constrained search at $\vec n^*$ are given by
\begin{equation}
  \ket{\Phi^*(\theta)}
  =
  \sqrt{\lambda}\ket{0,N,0}
  +\sqrt{1-\lambda}\,e^{i\theta}\ket{0,0,N}.
\end{equation}
This follows from two facts: first, the interaction $\hat W$ does not couple
states lying on the facet (a property of any two-body interaction), and second,
among all facet states, $\ket{0,N,0}$ and $\ket{0,0,N}$ have the lowest
interaction energy, which is specific to the Bose-Hubbard model.

The states $\ket{0,N,0}$ and $\ket{0,0,N}$ couple only to
$\ket{1,N-2,1}$ and $\ket{1,1,N-2}$, respectively, with matrix elements
$\braket{1,N-2,1|\hat W|0,N,0}
  =
  \braket{1,1,N-2|\hat W|0,0,N}
  =
  \frac{2}{3}\sqrt{N(N-1)}.$
Both off-facet states lie at the same distance from the facet,
$D^{(s)}(1,N-2,1)=D^{(s)}(1,1,N-2)=\sqrt{6}/2$. Substituting these ingredients
into Eq.~\eqref{eq:amazing_exforce_formula} yields
\begin{equation}
  \label{eq:hubbard_exforce}
  \begin{aligned}
    \ForcePrefactor[0,\lambda N,(1-\lambda)N]
    &=
    -2\left[
      \frac{
        \left(\frac{2}{3}\sqrt{N(N-1)}\right)^2
        \left(\lambda+(1-\lambda)\right)
      }{\frac{\sqrt{6}}{2}}
    \right]^{\!1/2}
    \\
    &=
    -\frac{4\cdot2^{1/4}\cdot3^{3/4}}{9}\sqrt{N(N-1)}.
  \end{aligned}
\end{equation}
This analysis applies for $N\ge6$; for $N=3$, the states
$\ket{1,N-2,1}$ and $\ket{1,1,N-2}$ coincide. Numerically, the prefactor is
approximately $1.205$. Notably, the resulting repulsion strength
$\ForcePrefactor[\vec n^*]$ is independent of the position $\lambda$ along the facet.

\begin{figure}[htb]
  \includegraphics[width=.75\columnwidth]{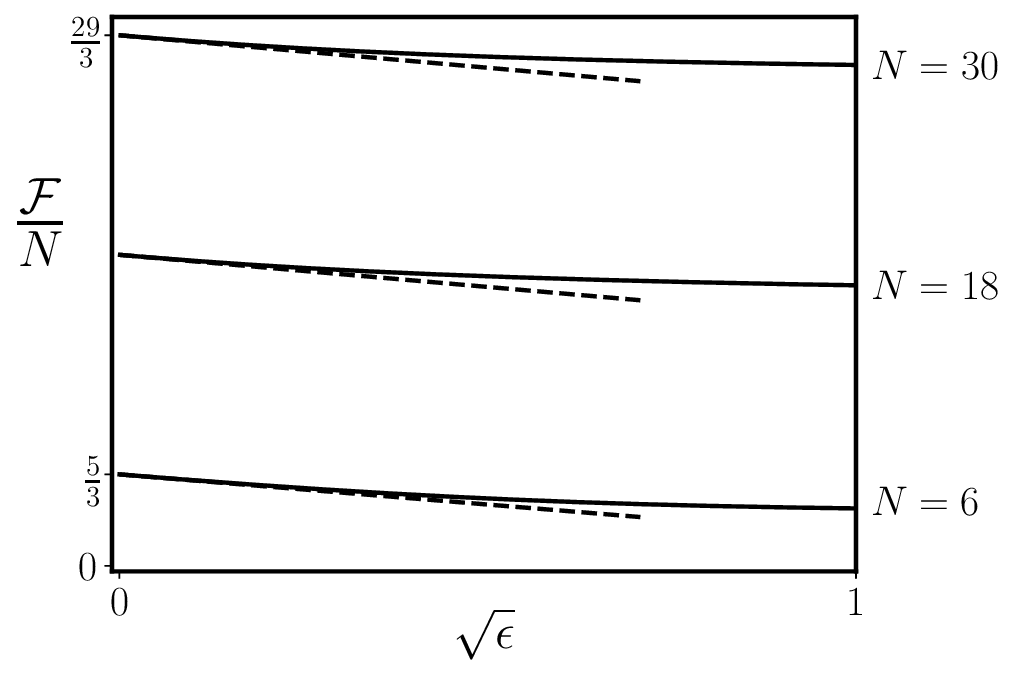}
  \caption{Comparison of the approximation
  \eqref{eq:hubbard_approximate_functional} (dashed lines) with exact numerical
  results (solid lines) for the universal functional in the Bose-Hubbard
  model \eqref{eq:hubbard_W} for $(d,N,P) = (3,N,0)$.}
  \label{fig:exforce_numer_d3NP0}
\end{figure}
To verify that Eq.~\eqref{eq:hubbard_exforce} indeed yields the correct repulsion
strength, we approximate the functional as
\begin{equation}
  \Fp[\vec n^*+\epsilon\vec\kappa^{(s)}]
  \approx
  \Fp[\vec n^*]
  +\sqrt{\epsilon}\,\ForcePrefactor[\vec n^*],
\end{equation}
and compare this approximation with numerical results. Choosing
$\vec n^*=(0,N/2,N/2)$, for which $\Fp[\vec n^*]=N(N-1)/3$, we obtain
\begin{equation}
  \label{eq:hubbard_approximate_functional}
  \frac{\Fp[\vec n^*+\epsilon\vec\kappa^{(s)}]}{N}
  \approx
  \frac{N-1}{3}
  -1.205\,\sqrt{\epsilon}\sqrt{\frac{N-1}{N}}.
\end{equation}
Figure~\ref{fig:exforce_numer_d3NP0} compares this expression to the exact
functional for $N=6,18,30$. The agreement of the slopes at $\epsilon=0$
demonstrates that Eq.~\eqref{eq:hubbard_exforce} provides the correct repulsion
strength at $\vec n^*$.

\begin{figure}
  \centering
  \includegraphics[width=1.03\columnwidth]{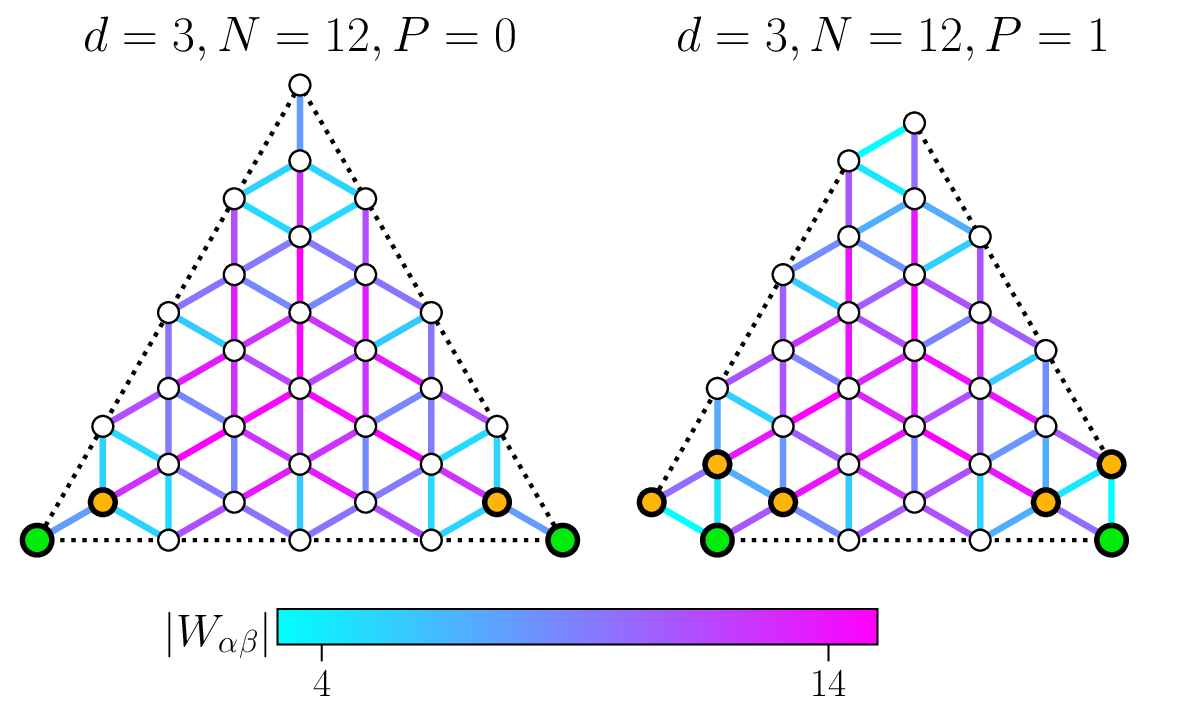}
  \caption{Graphical explanation of the stronger generalized BEC force for
  $P=1$. Two occupation number vectors $\vec{n}^{(\alpha)}, \vec{n}^{(\beta)}$ are connected if $W_{\alpha\beta}\neq0$,
  with color encoding of the coupling $|W_{\alpha\beta}|$ between their configuration states $|\vec{n}^{(\alpha)}\rangle, |\vec{n}^{(\beta)}\rangle$. On a given facet, the extremal states
  (green) couple to two off-facet states (orange) for $P=0$ and to five
  off-facet states for $P=1$. See text for details.}
  \label{fig:d12_hubbard_coupling}
\end{figure}

Finally, we remark that Eq.~\eqref{eq:amazing_exforce_formula} also explains the
larger generalized BEC force observed for $P=1$ compared to $P=0$ at fixed
$(d=3,N)$ when $N$ is a multiple of $3$ (see Fig.~\ref{fig:gradient_F_plots}).
For $P=1$, minimizing states on a facet still involve only the two extremal
states (cf.~Fig.~\ref{fig:d12_hubbard_coupling}), but these now couple to five
off-facet states, rather than two as in the $P=0$ case. Consequently, the sum in
Eq.~\eqref{eq:amazing_exforce_formula} contains five terms instead of two,
leading to a larger repulsion strength. This mechanism is illustrated in
Fig.~\ref{fig:d12_hubbard_coupling} for $N=12$.

To summarize this section, Eq.~\eqref{eq:amazing_exforce_formula} provides an explicit and universal
characterization of the leading boundary singularity of $\Fp$, separating
geometric information encoded in the facet distances $D^{(s)}(\nalpha)$ from
interaction effects entering through the matrix elements
$\braket{\nalpha|\hat W|\Phi^*(\vec n^*)}$. The resulting square-root divergence
of $\nabla_{\vec n}\Fp$ is therefore not a model-dependent artifact but a direct
consequence of the constrained-search structure near facets of $\dom(\Fp)$.
This perspective suggests a concrete route for constructing approximate
functionals that enforce the correct near-facet behavior by design, while
retaining flexibility in modeling the physically relevant interaction
contributions.

\section{Proof of concept: A complete illustrative example}\label{sec:ex}
In this section we present a self-contained, concrete analysis that
illustrates all essential ingredients developed in
Secs.~\ref{sec:domain}--\ref{sec:BECforce}. Beyond serving as an explicit
validation of our general framework, the example is designed to demonstrate
its practical relevance: it yields a compact representation of the exact
functional and provides a transparent route to accurate, physics-informed
functional approximations.

We consider the one-dimensional Bose-Hubbard model with $N=3$ bosons on
$d=3$ lattice sites and restrict to the total momentum sector $P=0$.
For this system, the functional $\Fp$ can in principle be obtained in closed
form, because it follows from the ground-state energies of the Hamiltonian
$\hat H(\hat t) = \hat{t}+\hat{W}$ that takes the form of a $4\times4$ matrix.
However, the resulting quartic expressions are cumbersome and offer little
insight. Instead, we proceed in a way that mirrors the general strategy of
this work and, crucially, produces an approximation for $\Fp$ that is both
simple and highly accurate when benchmarked against numerical calculations.

For clarity, we organize the discussion and derivation into three steps:
\begin{enumerate}[label=(\arabic*)]
  \item Identify the symmetry sector $\HN^{(P)}$ and determine the vertices spanning the domain $\dom(\Fp)$.
  \item Convert $\dom(\Fp)$ to a hyperplane (facet) representation; compute the matrix $T$ \eqref{eq:D_from_y_temp} and
        its kernel.
  \item Apply Eq.~\eqref{eq:general_F_form} and construct a controlled approximation,
        highlighting the role of the generalized BEC force.
\end{enumerate}
The approximation and its numerical validation are presented in
Sec.~\ref{sec:ex_step_3}.

\subsection{Identifying the symmetry sector and the domain of the functional}
The total $N$-particle Hilbert space $\HN$ is spanned by the ten configuration
states $\ket{3,0,0}, \ket{2,1,0}, \dotsb, \ket{0,0,3}$.
The $P=0$ symmetry sector $\HN^{(0)}$ is a four-dimensional subspace spanned by
\begin{eqnarray}
 && \ket{\vec n^{(1)}}\equiv \ket{3,0,0},\quad
  \ket{\vec n^{(2)}}\equiv \ket{0,3,0},\quad
  \ket{\vec n^{(3)}}\equiv \ket{0,0,3}, \nonumber \\
  && \hspace{2.93cm}\ket{\vec n^{(4)}}\equiv \ket{1,1,1}.
\end{eqnarray}
Their occupation number vectors $\vec{n}^{(\alpha)}, \alpha=1,\ldots, 4$, then span the domain of the universal functional $\mathcal{F}^{(0)}$ (see panel~(c) of Fig.~\ref{fig:domains_dNP}).
Restricted to $\mathcal{H}_{3}^{(0)}$ and expressed in this basis, the Bose-Hubbard
interaction operator $\hat W$ (Eq.~\eqref{eq:hubbard_W}) takes the form
\begin{equation}
\hat W = \begin{pmatrix}
2 & 0 & 0 & \frac{2\sqrt{6}}{3}\\
0 & 2 & 0 & \frac{2\sqrt{6}}{3}\\
0 & 0 & 2 & \frac{2\sqrt{6}}{3}\\
\frac{2\sqrt{6}}{3} & \frac{2\sqrt{6}}{3} & \frac{2\sqrt{6}}{3} &4\\
\end{pmatrix}.
\end{equation}
This explicit reduction already captures the first central theme of our work:
symmetries substantially simplify the structure of $\Fp$ by reducing both the
relevant Hilbert space and the geometry of the functional domain.

\subsection{Hyperplane (facet) representation and finding the kernel of $T$}
Following Sec.~\ref{sec:F_beyond_simplex}, we compute the facet constraints
$D^{(j)}$ and the matrix $T$ that encodes the distances of the vertices to the
facets. Since $\dom(\Fp)$ is a $2$-simplex (panel~(c) of
Fig.~\ref{fig:domains_dNP}), there are three facets and hence three constraints:
\begin{equation}
\begin{aligned}
  &D^{(1)}(\vec n) \equiv  \frac{1}{\sqrt{6}}(2n_0-n_1-n_2) + \frac{3}{\sqrt{6}} \ge 0\\
  &D^{(2)}(\vec n) \equiv  \frac{1}{\sqrt{6}}(2n_1-n_0-n_2) + \frac{3}{\sqrt{6}} \ge 0\\
  &D^{(3)}(\vec n) \equiv  \frac{1}{\sqrt{6}}(2n_2-n_0-n_1) + \frac{3}{\sqrt{6}} \ge 0,
\end{aligned}
\end{equation}
where the coefficients are chosen such that $|\vec \kappa^{(j)}|=1$ and
$\sum_{k=0}^{2}\kappa^{(j)}_k = 0$.
Recall that $T_{j\alpha}=D^{(j)}(\vec n^{(\alpha)})$ equals the distance from
the $\alpha$th vertex to the $j$th facet. Direct computation yields
\begin{equation}
  T = \frac{\sqrt{6}}{2} \begin{pmatrix}
  3 & 0 & 0 & 1\\
  0 & 3 & 0 & 1\\
  0 & 0 & 3 & 1\\
  \end{pmatrix}
\end{equation}
with pseudoinverse
\begin{equation}
T^+ = \frac{2}{\sqrt{6}}\frac{1}{36}\begin{pmatrix}
11 & -1 & -1\\
-1 & 11 & -1\\
-1 & -1 & 11\\
3 & 3 & 3
\end{pmatrix}.
\end{equation}
Since $T$ has rank $3$, its kernel is one-dimensional. One readily verifies
that it is spanned by $(1,1,1,-3)^\top$. Consequently, the radicands in
Eq.~\eqref{eq:general_F_form} admit the explicit parametrization
\begin{equation}
T^+\vec D(\vec n) + \vec x =
\begin{pmatrix}
\frac{n_0}{3} -\frac{1}{12}\\
\frac{n_1}{3} -\frac{1}{12}\\
\frac{n_2}{3} -\frac{1}{12}\\
\frac{1}{4} \\
\end{pmatrix}
+ \xi\begin{pmatrix}1 \\ 1 \\ 1 \\ -3\end{pmatrix}
 = \begin{pmatrix}
 \frac{n_0-z}{3}  \\
 \frac{n_1-z}{3}  \\
 \frac{n_2-z}{3}  \\
    z \\
 \end{pmatrix},
\end{equation}
where $z \equiv \frac{1}{4}-3\xi$.
This is the key structural simplification: the general representation
Eq.~\eqref{eq:general_F_form} reduces the remaining nontrivial dependence of
$\Fp$ to a \emph{single} scalar parameter $z$ associated with $\ker(T)$.
The ensuing approximation problem is therefore both low-dimensional and
geometrically constrained.

\subsection{Applying Eq.~\eqref{eq:general_F_form}}
\label{sec:ex_step_3}
Because $\ket{\vec n^{(1)}}$, $\ket{\vec n^{(2)}}$, and $\ket{\vec n^{(3)}}$
each couple only to $\ket{\vec n^{(4)}}$ (and not to one another), the
minimization over phases $\{\eta_\alpha\}$ in Eq.~\eqref{eq:general_F_form} is
immediate; one admissible choice is $\eta_4=-1$ and
$\eta_1=\eta_2=\eta_3=+1$.
With the phases fixed, $\Fp$ can be written as
\begin{equation}
  \label{eq:d3N3P0_functional_with_z}
\Fp[\vec n] = 2 + 2z - \frac{4\sqrt{6}}{3}\sum_{k=0}^2\sqrt{\frac{n_k-z}{3}}\sqrt{z},
\end{equation}
where the remaining task is the minimization over $z$.
In what follows, we denote by $\bar z(\vec n)$ the minimizing value at $\vec n$.

\paragraph{A constrained, physics-informed approximation for $\bar z$.}
From the radicands in the previous subsection we have
$0 \le z \le \min(n_0,n_1,n_2)\equiv z_{\text{max}}(\vec n)$.
A central point of this proof of concept is that, once the representation
Eq.~\eqref{eq:d3N3P0_functional_with_z} is available, constructing a useful
approximation for $\Fp$ reduces to proposing a plausible approximation for the
\emph{scalar} minimizer $\bar z(\vec n)$ consistent with (i) facet behavior and
(ii) regularity at symmetric points. We therefore make the simplifying ansatz
that $\bar z$ depends on $\vec n$ only through $z_{\text{max}}$ and construct a
function $\bar z(z_{\text{max}})$ that matches these constraints.

On the facet $n_0=0$ (hence $z_{\text{max}}=0$) we necessarily have $\bar z=0$.
Moreover, moving slightly away from that facet, Eq.~\eqref{eq:d3N3P0_functional_with_z}
becomes
\begin{equation}
  \label{eq:d3N3P0_functional_near_facet}
  \Fp[n_0, n_1, n_2] = 2  - \frac{4\sqrt{6}}{3}
  \left(\sqrt{\frac{n_1}{3}} + \sqrt{\frac{n_2}{3}}\right) \sqrt{z} + O(n_0),
\end{equation}
since $z\le n_0 \ll n_1,n_2$.
At this level, the minimum is attained by choosing $z=n_0$, suggesting the
boundary condition $\bar z'(0)=1$.
Importantly, the mechanism behind this behavior is precisely the one analyzed
in Sec.~\ref{sec:BECforce}: the $\sqrt{z}$-term induces the characteristic
repulsive tendency away from the facet, and therefore directly constrains the
admissible shape of $\bar z$ near $z_{\text{max}}=0$.

At the domain center, $z_{\text{max}}=1$ and $\vec n=(1,1,1)$, one finds from
Eq.~\eqref{eq:d3N3P0_functional_with_z}
\begin{equation}
  \Fp[1,1,1] = 2 + 2z - 4\sqrt{2}\sqrt{1-z}\sqrt{z},
\end{equation}
which is minimized at $z=\frac{1}{3}$.
We therefore impose the four conditions
\begin{equation}
  \label{eq:conditions_on_z}
  \begin{matrix}
    \bar z(0) = 0 & \bar z'(0) = 1\\
    \bar z(1) = \frac{1}{3} & \bar z'(1) = 0,
  \end{matrix}
\end{equation}
where $\bar z'(1)=0$ is chosen to ensure differentiability of the resulting
approximate functional at $\vec n=(1,1,1)$.

The simplest polynomial satisfying Eq.~\eqref{eq:conditions_on_z} is
\begin{equation}
  \label{eq:z_approx}
  \bar z^{\text{approx}} = z_{\text{max}} \left(\frac{z_{\text{max}}^2}{3} - z_{\text{max}} + 1\right),
\end{equation}
which yields an approximate functional (more precisely, an upper bound) when
inserted into Eq.~\eqref{eq:d3N3P0_functional_with_z}.

\paragraph{Validation against numerical minimization of $z$.}
\begin{figure}
  \centering
  \includegraphics[width=.82\columnwidth]{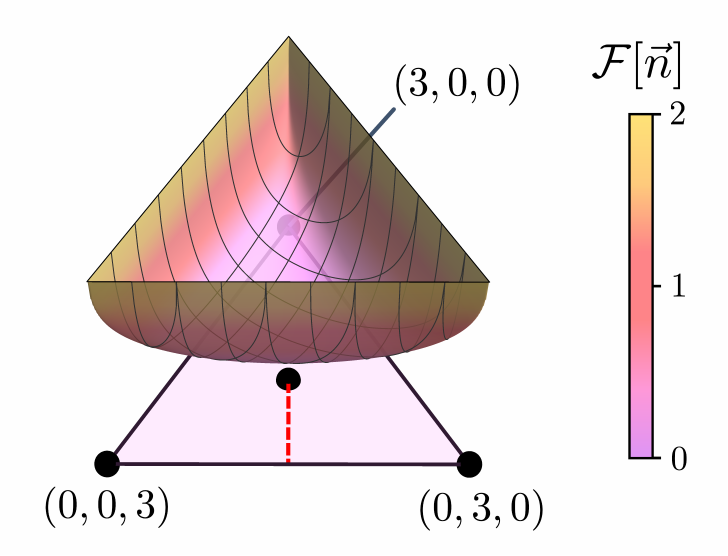}
  \caption{Exact functional for the Bose-Hubbard model with $(d, N, P) = (3,3,0)$.}
  \label{fig:hubbard_d3N3P0_functional}
\end{figure}
For Fig.~\ref{fig:hubbard_d3N3P0_functional}, we numerically minimize
Eq.~\eqref{eq:d3N3P0_functional_with_z} over $z$ and then plot the resulting exact
$\Fp$.
In Fig.~\ref{fig:hubbard_d3N3P0_approx_comparison}, the exact numerical
functional and minimizer $\bar z$ are compared to their approximated
counterparts obtained from Eq.~\eqref{eq:z_approx} along the straight path from
$(0,\frac{3}{2},\frac{3}{2})$ to $(1,1,1)$.
While $\bar z^{\text{approx}}$ exhibits small deviations from the exact
minimizer, the corresponding approximate functional is visually
indistinguishable from the exact $\Fp$ along that path.

\begin{figure}[htb]
  \label{fig:energy_functional_error}
	\centering
  \subfloat[functional error\label{fig:hubbard_d3N3P0_functional_diff}]{
    \includegraphics[width=.65\columnwidth]{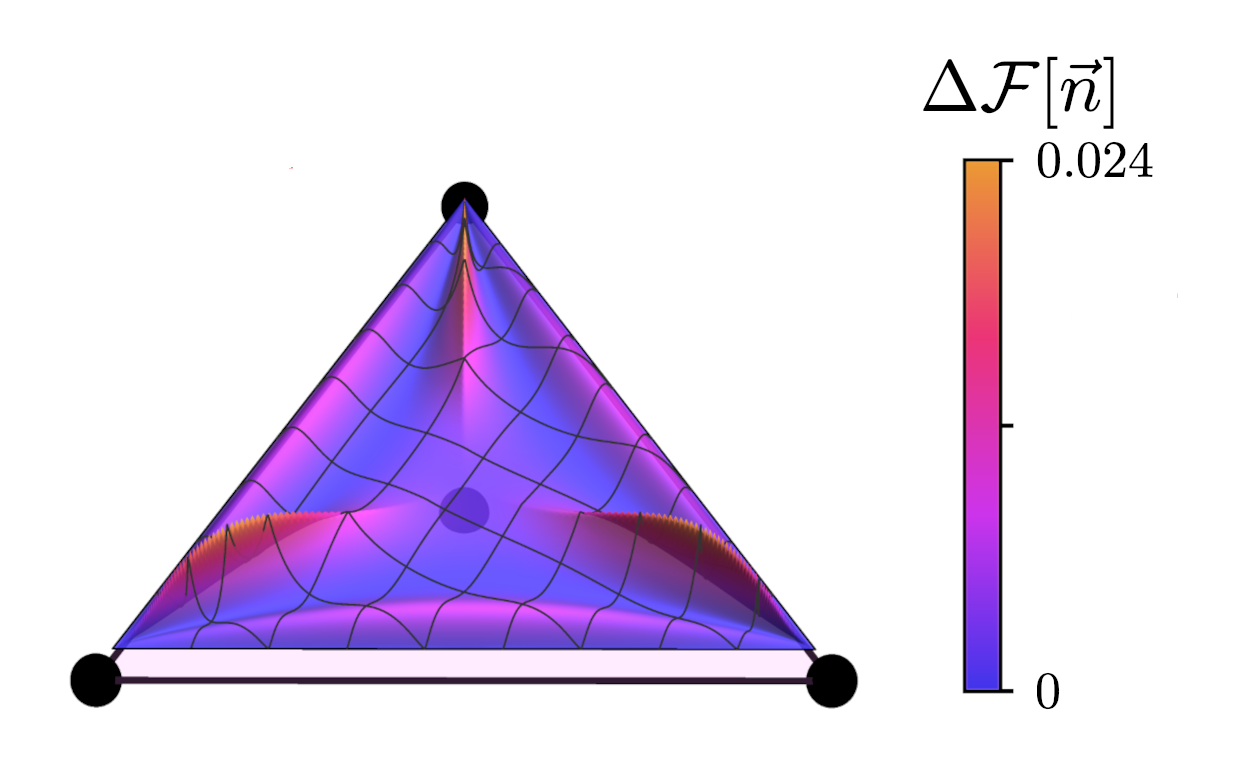}
  }

  \subfloat[exact ground state energy\label{fig:hubbard_d3N3P0_energy}]{
  	\includegraphics[width=.5\columnwidth]{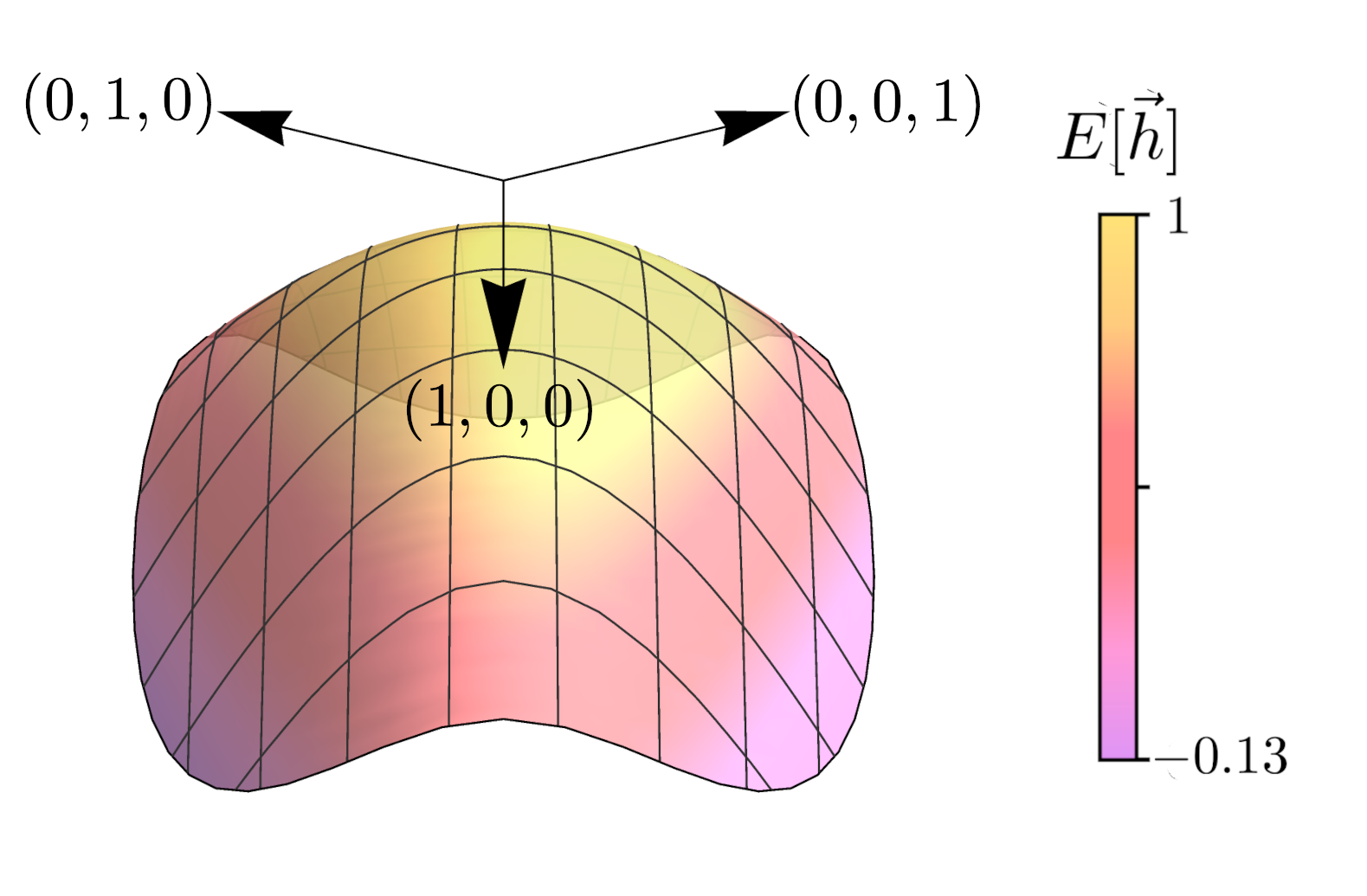}
  }
  \subfloat[ground state energy error\label{fig:hubbard_d3N3P0_energy_diff}]{
  	\includegraphics[width=.5\columnwidth]{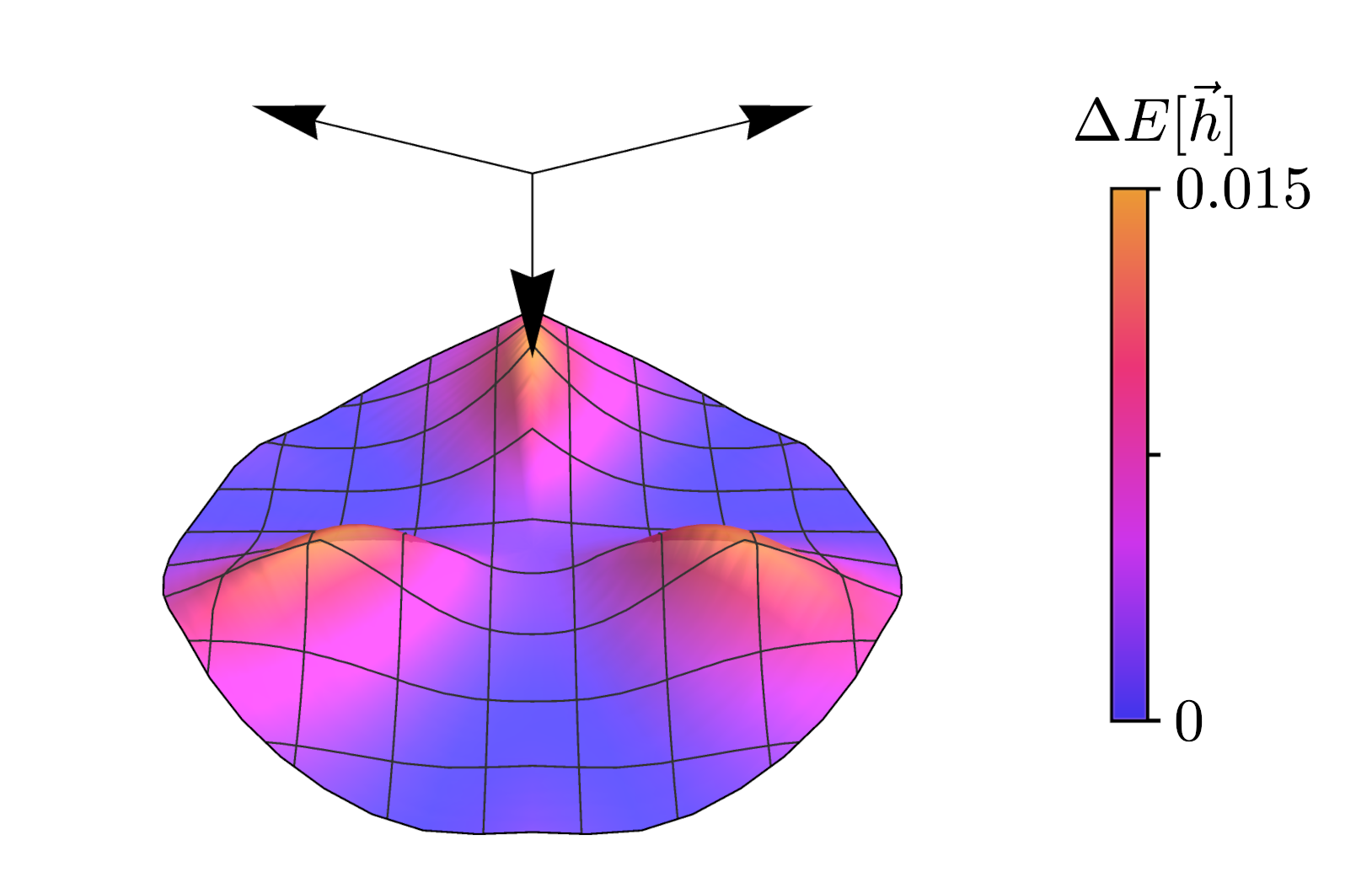}
  }
	\caption{The error of our functional approximation by plugging
	Eq.~\eqref{eq:z_approx} into Eq.~\eqref{eq:d3N3P0_functional_with_z} (top), along
	with the exact ground state energy (bottom left) and 
  the respective error (bottom right). The arrows in the energy plots indicate
  directions in the $3$-dimensional $\vec h$-space.
	}
\end{figure}

The pointwise error of the approximate functional over the full domain is
shown in Fig.~\ref{fig:hubbard_d3N3P0_functional_diff}.
The largest deviations occur near the line segments connecting $(1,1,1)$ to
$(3,0,0)$, $(0,3,0)$, and $(0,0,3)$, and in their vicinity.
Nevertheless, we find numerically that the maximum error is $0.024$, which is
only about $1\%$ of $\max \Fp(\vec n)-\min \Fp(\vec n)$ on the domain (see
Fig.~\ref{fig:hubbard_d3N3P0_functional}).

\begin{figure}[htb]
  \centering
  \includegraphics[width=.95\columnwidth]{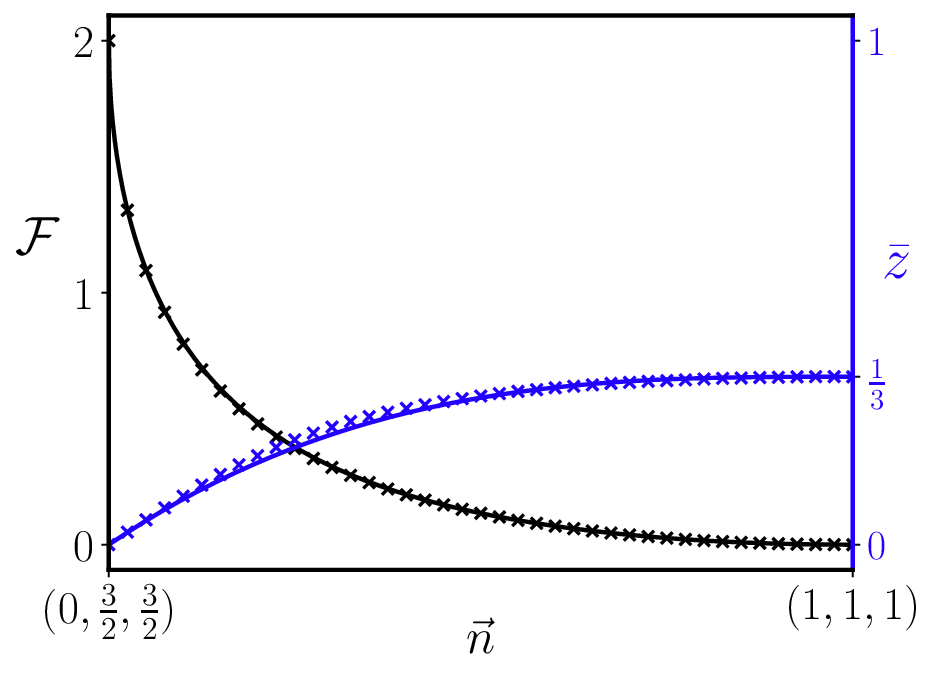}
  \caption{
    Comparison between approximated $\Fp$ and $\bar z$ (solid curves) and
    the numerical values (cross markers) along the dotted line segment in
    Fig.~\ref{fig:hubbard_d3N3P0_functional}.
  }
  \label{fig:hubbard_d3N3P0_approx_comparison}
\end{figure}

\paragraph{Impact on energies: accuracy of the induced variational problem.}
Since $\Fp^{\text{approx}}$ stays close to the exact functional throughout the
domain, the corresponding approximate ground-state energy
\begin{equation}
E_{\text{approx}}[\vec t]=\min_{\vec n}\bigl(\vec t\cdot \vec n +
\Fp^{\text{approx}}[\vec n]\bigr)
\end{equation}
is expected to be accurate as well.
Figure~\ref{fig:hubbard_d3N3P0_energy} displays the exact energy for the family
of one-particle Hamiltonians
\begin{equation}
\vec t(r,\theta) =
\begin{pmatrix}
  \frac{-1}{\sqrt{6}} & -\frac{1}{\sqrt{2}}\\
  \frac{-1}{\sqrt{6}} & \frac{1}{\sqrt{2}}\\
  \frac{2}{\sqrt{6}} & 0
  \end{pmatrix}
  \begin{pmatrix}r\cos \theta\\ r\sin \theta\end{pmatrix}
  + \frac{1}{3}\begin{pmatrix}1\\1\\1\end{pmatrix},
\end{equation}
with $r\in[0,1]$ and $\theta\in[0,2\pi]$.
(The last term merely adds a constant to the Hamiltonian and does not affect
the physics.)
Geometrically, this parameterizes a disk in the affine space
$\{\vec t\mid t_0+t_1+t_2=1\}\subset \mathbb{R}^3$.
The error
$\Delta E[\vec t]\equiv E_{\text{approx}}[\vec t]-E[\vec t]$
is shown in Fig.~\ref{fig:hubbard_d3N3P0_energy_diff}; over the displayed
parameter range, the approximate energy is again accurate to about $1\%$ of
$\max E[\vec h]-\min E[\vec h]$ on the disk.

To conclude, this example serves as a proof of concept that the fruitful geometric
structure of $\dom(\Fp)$ and the BEC force can be exploited to construct sophisticated functional
approximations in a controlled and transparent manner.

\section{Summary and Conclusion}
\label{sec:summary_conclusion}

In this work, we have developed a symmetry-adapted reduced density matrix
functional theory (RDMFT) framework for interacting bosonic lattice systems.
Building on the constrained-search formulation of RDMFT, we demonstrated that
translational invariance, together with a fixed interaction $\hat W$, leads to a
decisive simplification of the functional description: the basic functional
variable reduces unambiguously from the full one-particle reduced density matrix
$\hat\gamma$ to the momentum occupation number vector $\vec n$. As a result, the
ground-state energy of translationally invariant lattice bosons can be described
by a universal interaction functional defined solely on this reduced set of
variables. This formulation is directly relevant for ultracold atomic gases,
where momentum distributions are experimentally accessible and
Bose-Einstein condensation admits a natural characterization within RDMFT
through the Penrose-Onsager criterion.

A central result of the paper is that the structure of this universal functional
is governed by the geometry of its domain, which is completely determined by
one-body $N$-representability. We showed that the admissible set of momentum
occupation numbers forms a reduced, symmetry-adapted convex polytope. By
expressing the functional in coordinates adapted to the facets of this polytope,
we derived a general form in which the dependence on the occupation numbers is
dictated primarily by geometric considerations rather than by the detailed form
of the interaction.

This geometric perspective has direct physical consequences. We demonstrated
that the geometry of the representability domain enforces a universal boundary
behavior of the functional: as the occupation number vector approaches any facet
of the domain, the gradient of the functional diverges repulsively. This
generalized BEC force extends earlier observations near the condensation vertex
to arbitrary boundaries of the domain. Going beyond qualitative arguments, we
derived a closed analytical expression for the corresponding repulsion strength. 
The resulting boundary behavior is therefore not a model-dependent
feature but a structural consequence of the constrained-search formulation and
of $N$-representability, and is expected to arise generically in interacting
bosonic quantum systems.

Beyond its conceptual implications, the geometric form of the functional
provides concrete guidance for the construction of approximations. The exact
near-boundary behavior imposes nontrivial constraints on admissible functionals,
while at the same time suggesting a hierarchy of geometry-informed approximation
schemes. We illustrated this strategy analytically for a few-site lattice system,
demonstrating how exact geometric input can be combined with
interaction-specific information in a controlled and transparent manner. This
example serves as a proof of concept for a broader approximation program.

Taken together, our results show that spatial symmetry,
$N$-representability, and the geometry of quantum states are intrinsically linked,
as anticipated in Fig.~\ref{fig:overview}, and jointly determine the structure of
bosonic reduced density matrix functionals. While our explicit constructions were
carried out for one-dimensional lattice systems, the generality of the geometric
framework makes it clear that analogous results can be obtained for higher
spatial dimensions and for bosons carrying additional internal degrees of
freedom, such as spin. Moreover, several of the key insights derived here extend,
at least qualitatively, to ensemble formulations, reflecting the general
principle that functional theories are defined by two fundamental ingredients: a
scope, specifying the class of physical systems under consideration, and a
variational principle, such as the Rayleigh-Ritz principle for ground states or
its ensemble generalizations for excited-state theories.

At a more fundamental level, our results suggest a shift in perspective on the
choice of basic variables in functional theories. The decisive role played by the
distances of $\vec n$ to the facets of the representability domain indicates that
it is often natural to reparameterize the functional not in terms of the density
matrix or occupation numbers themselves, but in terms of their geometric
distances to the boundary of the admissible set. Making this structure explicit
provides direct physical insight, constrains admissible functional forms, and
offers a principled starting point for systematic approximations. In this way,
the present work closes a conceptual gap in bosonic RDMFT and lays the foundation
for a geometry-driven approach to predictive functional theories of strongly
correlated bosonic systems.

\begin{acknowledgments}
We are grateful to Markus Penz for valuable feedback on the manuscript. We acknowledge financial support from the German Research Foundation (Grant SCHI 1476/1-1). 
\end{acknowledgments}

\appendix

\section{Derivation of the Generalized BEC Force}
\label{app:BECforce}

Here we present the derivation of the generalized BEC force without the simplex
assumption. We adopt the notation that $I^{(j)}$ denotes all indices $\alpha$
for which the occupation number vector $\nalpha$ satisfies
$D^{(j)}(\nalpha)=0$. In other words, $I^{(j)}$ is the set of indices of all
occupation number vectors on facet~$j$. The intuition is as follows: whenever a
density $\vec n$ approaches one of the facets, say
$D^{(s)}(\vec n)=\epsilon\ll 1$, the wave function must assume a form such that
the coefficients of configuration states $\ket{\nalpha}$ not on the facet
approach zero at a rate proportional to $\sqrt{\epsilon}$. In this regime, the
zeroth-order contribution to the functional arises from the configuration
states on the facet, to which a correction of order $\sqrt{\epsilon}$ from
facet--off-facet interactions is added. This latter term is the origin of the
generalized BEC force.

Let $\vec n^*$ be a point on facet~$s$, i.e., $D^{(s)}(\vec n^*)=0$. Let
$\ket{\Phi(\epsilon)}$, with $\epsilon\in[0,\epsilon')$, be a curve such that
$\ket{\Phi(\epsilon)}\mapsto \vec n^* + \epsilon \vec\kappa^{(s)}$, and such
that $\ket{\Phi(\epsilon)}$ is a minimizer of the constrained search for each
$\epsilon$. That is,
\begin{equation}
  \label{eq:functional_near_facet}
  \Fp[\vec n^* + \epsilon \vec \kappa^{(s)}]
  = \braket{\Phi(\epsilon)|\hat W|\Phi(\epsilon)}.
\end{equation}
Write $\ket{\Phi(\epsilon)}=\sum_{\alpha}c_\alpha(\epsilon)\ket{\nalpha}$. We
assume that the coefficients $c_\alpha(\epsilon)$ take the following form:
\begin{equation}
  \label{eq:bec_force_coeff_ansatz}
  \begin{cases}
    c_\alpha(\epsilon)=p_\alpha + \epsilon q_\alpha + O(\epsilon^2),
    & \alpha \in I^{(s)},\\
    c_\alpha(\epsilon)=\sqrt{\epsilon}\,r_\alpha + O(\epsilon),
    & \alpha \notin I^{(s)}.
  \end{cases}
\end{equation}
Equation~\eqref{eq:bec_force_coeff_ansatz} requires some justification. Since
$\braket{\Phi(\epsilon)|\hat n_k|\Phi(\epsilon)}$ depends linearly on each
$|c_\alpha(\epsilon)|^2$, and only the `off-facet' states, i.e.,
$\ket{\nalpha}$ with $\alpha\notin I^{(s)}$, can contribute to the deviation
$\epsilon\vec \kappa^{(s)}$ from the facet, we must have
$c_\alpha(\epsilon)\sim\sqrt{\epsilon}$ for $\alpha\notin I^{(s)}$, up to
higher-order corrections in $\epsilon$. On the other hand, it is not
immediately clear that the `on-facet' coefficients, i.e.,
$c_\alpha(\epsilon)$ with $\alpha\in I^{(s)}$, should not contain terms
proportional to $\sqrt{\epsilon}$. Indeed, if $c_\alpha(0)=0$, it is
conceivable that $c_\alpha(\epsilon)\sim\sqrt{\epsilon}$ even if
$\alpha\in I^{(s)}$. However, the vanishing of $c_\alpha(0)$ implies that the
state $\ket{\nalpha}$ is decoupled from other states on the facet, so
$c_\alpha$ does not contribute to the generalized BEC force.

Plugging Eq.~\eqref{eq:bec_force_coeff_ansatz} into
Eq.~\eqref{eq:functional_near_facet} yields
\begin{eqnarray}
    \Fp[\vec n^* + \epsilon \vec \kappa^{(s)}]
    &=& \sum_{\alpha,\beta \in I^{(s)}}
    p_\alpha^* p_\beta \braket{\nalpha|\hat W|\vec n^{(\beta)}}\\
    &&\hspace{-4em}\quad + 2\sqrt{\epsilon}\,\mathrm{Re}
    \sum_{\alpha\in I^{(s)},\, \beta\notin I^{(s)}}
    p_\alpha^* r_\beta
    \braket{\vec n^{(\alpha)}|\hat W | \vec n^{(\beta)}} + O(\epsilon).\nonumber 
\end{eqnarray}
The coefficients $p_\alpha$ must be chosen so as to minimize the first term, and
the result is simply $\Fp[\vec n^*]$. Similarly, the coefficients $r_\alpha$
are fixed by minimizing the second term. Observe that the phases are not
frustrated: there is always a choice of phases that minimizes the expression.
Thus,
\begin{equation}
  \label{eq:bec_force_intermediate_form}
  \begin{aligned}
    \frac{\Fp[\vec n^* + \epsilon\vec \kappa^{(s)}] - \Fp[\vec n^*]}{\sqrt{\epsilon}}
    &= - 2 \sum_{\beta \notin I^{(s)}} |r_\beta|\,
    \Big|\sum_{\alpha\in I^{(s)}}
    p_\alpha^*\braket{\vec n^{(\alpha)}|\hat W|\vec n^{(\beta)}}\Big|\\
    &= -2\sum_{\beta \notin I^{(s)}} |r_\beta|\,
    \Big|\braket{\Phi^*| \hat W | \vec n^{(\beta)}}\Big|,
  \end{aligned}
\end{equation}
where we have defined
$\ket{\Phi^*}\equiv \ket{\Phi(0)}=\sum_{\alpha\in I^{(s)}} p_\alpha\ket{\nalpha}$.
The quantities $|r_\beta|$ cannot be chosen arbitrarily: the condition
$\ket{\Phi(\epsilon)}\mapsto \vec n^* + \epsilon\kappa^{(s)}$ implies
\begin{equation}
\sum_{\alpha\in I^{(s)}} \nalpha
\big(|p_\alpha|^2 + 2\epsilon\,\mathrm{Re}\,p_\alpha^* q_\alpha\big)
+ \epsilon\sum_{\alpha\notin I^{(s)}} \nalpha |r_\alpha|^2
= \vec n^* + \epsilon \vec \kappa^{(s)},
\end{equation}
and therefore
\begin{equation}
2\mathrm{Re}\sum_{\alpha\in I^{(s)}} \vec n^{(\alpha)} p_\alpha^* q_\alpha
+ \sum_{\alpha\notin I^{(s)}} \vec n^{(\alpha)} |r_\alpha|^2
= \vec \kappa^{(s)}.
\end{equation}
Taking the dot product with $\vec \kappa^{(s)}$ and using
$\vec \kappa^{(s)}\cdot\nalpha=-\mu^{(s)}$ for $\alpha\in I^{(s)}$, we obtain
\begin{equation}
-2\mu^{(s)}\,\mathrm{Re}\sum_{\alpha\in I^{(s)}} p_\alpha^*q_\alpha
+ \sum_{\alpha\notin I^{(s)}} |r_\alpha|^2\,
\vec\kappa^{(s)}\cdot \nalpha
= |\vec \kappa^{(s)}|^2 = 1.
\end{equation}
(Recall that we assumed $\vec\kappa^{(s)}\cdot \vec\kappa^{(s)}=1$ at the
beginning of Sec.~\ref{sec:BECforce}.)
Normalization, $\braket{\Phi(\epsilon)|\Phi(\epsilon)}=1$, implies
\begin{equation}
2\mathrm{Re}\sum_{\alpha\in I^{(s)}}p_\alpha^*q_\alpha
= -\sum_{\alpha\notin I^{(s)}}|r_\alpha|^2,
\end{equation}
which leads to
\begin{equation}
  \label{eq:bec_force_coeff_constraint}
  \sum_{\alpha\notin I^{(s)}}
  |r_\alpha|^2 D^{(s)}(\nalpha) = 1.
\end{equation}
That is, the final line of Eq.~\eqref{eq:bec_force_intermediate_form} must be
minimized subject to the constraint
\eqref{eq:bec_force_coeff_constraint}. This constrained minimization can be
carried out explicitly and yields
\begin{eqnarray}
  \ForcePrefactor[\vec n^*]
  &=& \frac{\Fp[\vec n^*+\epsilon\vec\kappa^{(s)}] - \Fp[\vec n^*]}{\sqrt{\epsilon}}\\
  &=& -2 \left[
  \sum\limits_{\alpha:\,D^{(s)}(\nalpha)>0}
  \frac{\left|\braket{\nalpha|\hat W|\Phi^*}\right|^2}
       {D^{(s)}(\vec n^{(\alpha)})}
  \right]^{1/2}. \nonumber 
\end{eqnarray}

\bibliography{Refs4,Refs_extra}

\newpage


\end{document}